\documentclass{aa}
\pdfoutput=1
\usepackage{xspace}
\usepackage{natbib}
\usepackage{graphicx}
\usepackage{amssymb}
\usepackage{amsmath}
\usepackage{url}
\usepackage{txfonts}
\usepackage{rotating}

\usepackage{color}

\graphicspath{{./figs/}}

\begin{document}

\title{Long term variability of Cygnus X-1}
\subtitle{V. State definitions with all sky monitors}

\titlerunning{Long term variability of Cygnus X-1: V.  
State definitions with all sky monitors }

\author{V. Grinberg et al.}

\author{\mbox{V.~Grinberg\inst{1}} \and
  \mbox{N.~Hell\inst{1,2}} \and
\mbox{K.~Pottschmidt\inst{3,4}} \and
  \mbox{M.~B\"ock\inst{5}} \and
  \mbox{M.~A.~Nowak\inst{6}} \and
 \mbox{J.~Rodriguez\inst{7}} \and  
 \mbox{A.~Bodaghee\inst{8}}\and
  \mbox{M.~Cadolle Bel\inst{9}}\and
  \mbox{G.~L.~Case\inst{10}} \and
 \mbox{M.~Hanke\inst{1}} \and
  \mbox{M.~K\"uhnel\inst{1}} \and
  \mbox{S.~B.~Markoff\inst{11}} \and
 \mbox{G.~G.~Pooley\inst{12}} \and
  \mbox{R.~E.~Rothschild\inst{13}} \and
  \mbox{J.~A.~Tomsick\inst{8}} \and
    \mbox{C.~A.~Wilson-Hodge\inst{14}} \and
  \mbox{J.~Wilms\inst{1}}
}
\offprints{V.~Grinberg,\\ e-mail: {victoria.grinberg@fau.de}}
\institute{
Dr.\ Karl-Remeis-Sternwarte  and Erlangen Centre for
Astroparticle Physics, Friedrich Alexander Universit\"at
Erlangen-N\"urnberg, Sternwartstr.~7, 96049 Bamberg, Germany
\and
Lawrence Livermore National Laboratory, 7000 East Ave.,
Livermore, CA 94550, USA
\and
CRESST, University of Maryland Baltimore County, 1000 Hilltop Circle,
Baltimore, MD 21250, USA 
\and
NASA Goddard Space Flight Center, Astrophysics Science Division, Code
661, Greenbelt, MD 20771, USA 
\and
Max-Planck-Institut f\"ur Radioastronomie, Auf dem H\"ugel 69, 53121
Bonn, Germany
\and
MIT-CXC, NE80-6077, 77 Mass.\ Ave., Cambridge, MA 02139, USA
\and
Laboratoire AIM, UMR 7158, CEA/DSM - CNRS - Universit\'{e} Paris Diderot, 
IRFU/SAp, F-91191 Gif-sur-Yvette, France
\and
Space Sciences Laboratory, 7 Gauss Way, University of California,
Berkeley, CA 94720-7450, USA  
\and
European Space Astronomy Centre (ESA/ESAC), Science Operations
Department, PO Box 78, 28691 Villa\~{n}ueva de la Can\~{a}da, Madrid, Spain 
\and 
Department of Physics, La Sierra University, Riverside, CA, USA 92515
\and
Astronomical Institute ``Anton Pannekoek'', University of Amsterdam,
Kruislaan 403, Amsterdam, 1098 SJ, The Netherlands 
\and
Astrophysics, Cavendish Laboratory, University of Cambridge, CB3 0HE, UK
\and
Center for Astrophysics and Space Sciences, University of California
at San Diego, La Jolla, 9500 Gilman Drive, CA 92093-0424, USA 
\and 
ZP 12, NASA Marshall Space Flight Center, Huntsville, AL 35812, USA
}
\date {Received: --- / Accepted: ---}

\abstract{We present a scheme to determine the spectral state of the
  canonical black hole Cyg~X-1 using data from previous and current
  X-ray all sky monitors (RXTE-ASM, Swift-BAT, MAXI, and Fermi-GBM).
  State determinations of the hard/intermediate and soft state agree
  to better than 10\% between different monitors, facilitating the
  determination of the state and its context for any observation of
  the source, potentially over the lifetimes of different individual
  monitors. A separation of the hard and the intermediate state, which
  strongly differ in their spectral shape and short-term timing
  behavior, is only possible when monitor data in the soft X-rays
  ($<$5\,keV) are available. A statistical analysis of the states
  confirms the different activity patterns of the source (e.g., months
  to years long hard state periods or phases during which numerous
  transitions occurs). It also shows the hard and soft states to be
  stable, with the probability of Cyg X-1 remaining in a given state
  for at least one week to be larger than 85\% for the hard state and
  larger than 75\%, for the soft state.  Intermediate states are short
  lived, with a probability of 50\% that the source leaves the
  intermediate state within three days. A reliable detection of these
  potentially short-lived events is only possible with monitor data
  with a time resolution of better than 1\,d. }


\maketitle

\section{Introduction}

Accreting galactic black hole binaries (BHBs) show two main spectral
states: a soft state with a thermal X-ray spectrum dominated by an
accretion disk and a hard state with a power law spectrum with a
photon index $\Gamma\sim1.7$. The intermediate or transitional
state can be subdivided into a hard intermediate state and a soft
intermediate state \citep{Belloni_2010a}. Transient BHBs also show a
quiescent state \citep{McClintock_Remillard_2006a_book}. All states
show distinct spectral and timing properties in the X-rays. Radio
emission is detected in the hard and hard intermediate states and is
strongly suppressed in the soft states \citep[e.g.,][]{Fender_2009a}.

In a hardness intensity diagram (HID) transient BHBs move on a clear
trajectory, the so-called \textsf{q}-track \citep{Fender_2004a}.
Different states correspond to different parts of the track. Of
special interest is the so-called jet-line that roughly coincides with
the transition from hard intermediate to soft intermediate state and
is associated with radio ejection events \citep{Fender_2009a}. A
\textsf{q}-track-like behavior on HIDs or on their generalization, the
disk-fraction luminosity diagrams, has been found in different
accreting sources such as neutron star X-ray binaries
\citep[e.g.,][\object{Aql X-1}]{Maitra_2004a}, dwarf novae
\citep[][\object{SS~Cyg}]{Koerding_2008a}, or AGN
\citep{Koerding_2006a}. It is therefore likely to reflect basic
accretion/ejection physics inherent to a wide range of accreting
objects.

\object{Cygnus~X-1} is a key source for understanding accretion and
ejection processes and their connection in BHBs. It is persistent,
relatively nearby with a radio parallax distance of
$1.86^{+0.12}_{-0.11}$\,kpc \citep[][consistent with X-ray dust
scattering halo estimates of \citealt{Xiang_2011a}]{Reid_2011a},
bright (above $\sim$100\,mCrab in the 1.5--12\,keV band in the hard
state), and often undergoes (failed) state transitions
\citep[e.g.,][]{Pottschmidt_2003b}, which are thought to be connected
to changes in its radio jet
\citep{Fender_2004a,Fender_2006a,Wilms_2007a}. The black hole is in a
5.6\,d orbit around its donor star, HDE 226868 \citep[][and references
therein]{Brocksopp_1999a}. The orbital modulation is also detected in
radio \citep[e.g.,][]{Pooley_1999a} and, due to modulation of the soft
X-ray flux by absorption in the donor's stellar wind, in the X-rays
\citep{Balucinska_Church_2000a,Poutanen_2008a}. Additionally,
\mbox{Cyg\,X-1}\xspace shows a superorbital period of 150\,d
\citep{Brocksopp_1999b,Benlloch_2004a,Poutanen_2008a}, although that
superorbital variability seems to be unstable and has recently been
reported to have doubled to 300\,d \citep{Zdziarski_2011a}. In the
hard state, radio jets have been observed
\citep[e.g.,][]{Stirling_2001a}.  The emisssion above $\sim$400\,keV
\mbox{Cyg\,X-1}\xspace is strongly polarized
\citep{Laurent_2011a,Jourdain_2012a}.

As a persistent source \mbox{Cyg\,X-1}\xspace does not cover the full
\textsf{q}-track: its bolometric luminosity changes only by a factor
of $\sim$3--4 between the states \citep[][and references
therein]{Wilms_2006a} and its spectrum is never fully
disk-dominated. The frequent state transitions \citep[sometimes very
fast -- within hours, see,][]{Boeck_2011a} mean that the source often
crosses the jet line.  Since these state changes are thought to be
associated with significant changes in the accretion flow geometry and
energetics, a knowledge of the source state is crucial for the
interpretation of all (multiwavelength) observations of
\mbox{Cyg\,X-1}\xspace and its donor star. A typical example is the
study of the stellar wind of HDE 226868, which during soft states is
strongly photoionized by the radiation from the vicinity of the black
hole \citep{Gies_2008a}.

In the past decade, state information was readily available using the
All Sky Monitor on the Rossi X-ray Timing Explorer (RXTE-ASM) and
regular pointed monitoring observations. Here, various state
definitions exist, which use, e.g., measured count rates and/or colors
\citep[e.g.,][]{remillard:05a,Gies_2008a}, or sophisticated mapping
between these measurements and spectral parameters
\citep[e.g.,][]{Ibragimov_2007a,Zdziarski_2011a}. The former
prescription is easy to use, but is very instrument specific and
cannot be translated easily to other X-ray all sky monitors.  The
latter approach requires a sophisticated knowledge of the
instrumentation of all sky monitors as well as of the detailed
spectral modeling. Furthermore, the previously used state definitions
are all slightly inconsistent with each other. In this work we
introduce a novel approach to classify states of
\mbox{Cyg\,X-1}\xspace using the all sky monitors RXTE-ASM, MAXI,
Swift-BAT and Femi-GBM based on 16 years of pointed RXTE
observations. Our aim is to find an easy-to-use prescription for the
determination of states that is as consistent as possible between
these instruments in order to facilitate long-term studies that exceed
in duration the lifetime of individual monitors. We start with a
description of our data reduction approach in Sect.~\ref{sec:data}.
Section~\ref{sec:id_states} comprises the actual state mapping from
pointed RXTE observations to RXTE-ASM, Swift-BAT, MAXI and Fermi-GBM,
including a discussion of the precision of state determinations
attainable with these instruments. We summarize and discuss our
results in the light of the statistics of the state behavior of
\mbox{Cyg\,X-1}\xspace in Sects.~\ref{sec:states}
and~\ref{sec:summary}.

\section{Observations and Data Analysis}\label{sec:data}
\subsection{ASM data}\label{sec:asm_data}

\begin{figure*}
\includegraphics[width=\textwidth]{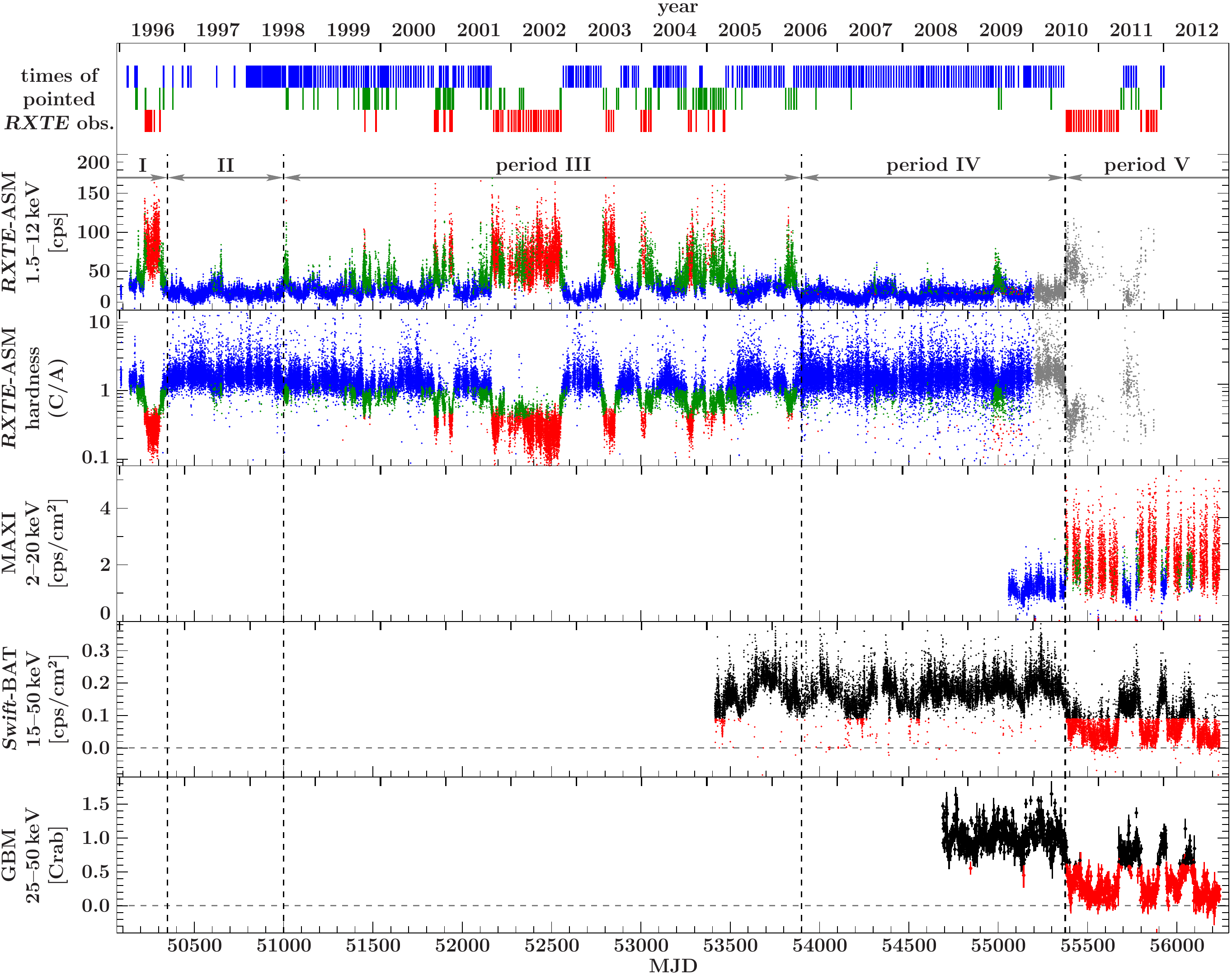}
\caption{Pointed RXTE observations and all light curves (RXTE-ASM,
  MAXI, Swift-BAT and GBM) of \mbox{Cyg\,X-1}\xspace used in this
  analysis. ASM, MAXI and BAT data are shown in the highest available
  resolution, GBM are binned daily. ASM hardness is calculated by
  dividing count rates in band C (5.0--12\,keV) by count rates in band
  A (1.5--3.0\,keV).  Vertical dashed lines and horizontal arrows
  represent periods of different source activity patterns. Blue,
  green, and red colors represent states of individual measurements
  classified using the respective classification for the different
  instruments as introduced in Sect.~\ref{sec:id_states}: blue
  represents the hard state, green the intermediate state and red the
  soft state.  ASM data after MJD~55200 (shown in grey) are affected
  by intrumental decline. Hard and intermediate states cannot be
  separated in BAT and GBM; BAT and GBM data corresponding to these
  periods of hard or intermediate states are therefore shown shown in
  black.}\label{fig:monster}
\end{figure*}

The RXTE-ASM instrument consisted of three Scanning Shadow Cameras
(SSCs), in which a position sensitive proportional counter was
illuminated through a slit mask \citep{Levine_1996a}. A typical source
was observed at randomly distributed times 5 to 10 times a day in
three energy bands roughly corresponding to 1.5--3.0\,keV (band~A),
3.0--5.0\,keV (band~B), and 5.0--12\,keV (band~C)
\citep{Levine_1996a}.

\begin{figure*}
\includegraphics[width=\columnwidth]{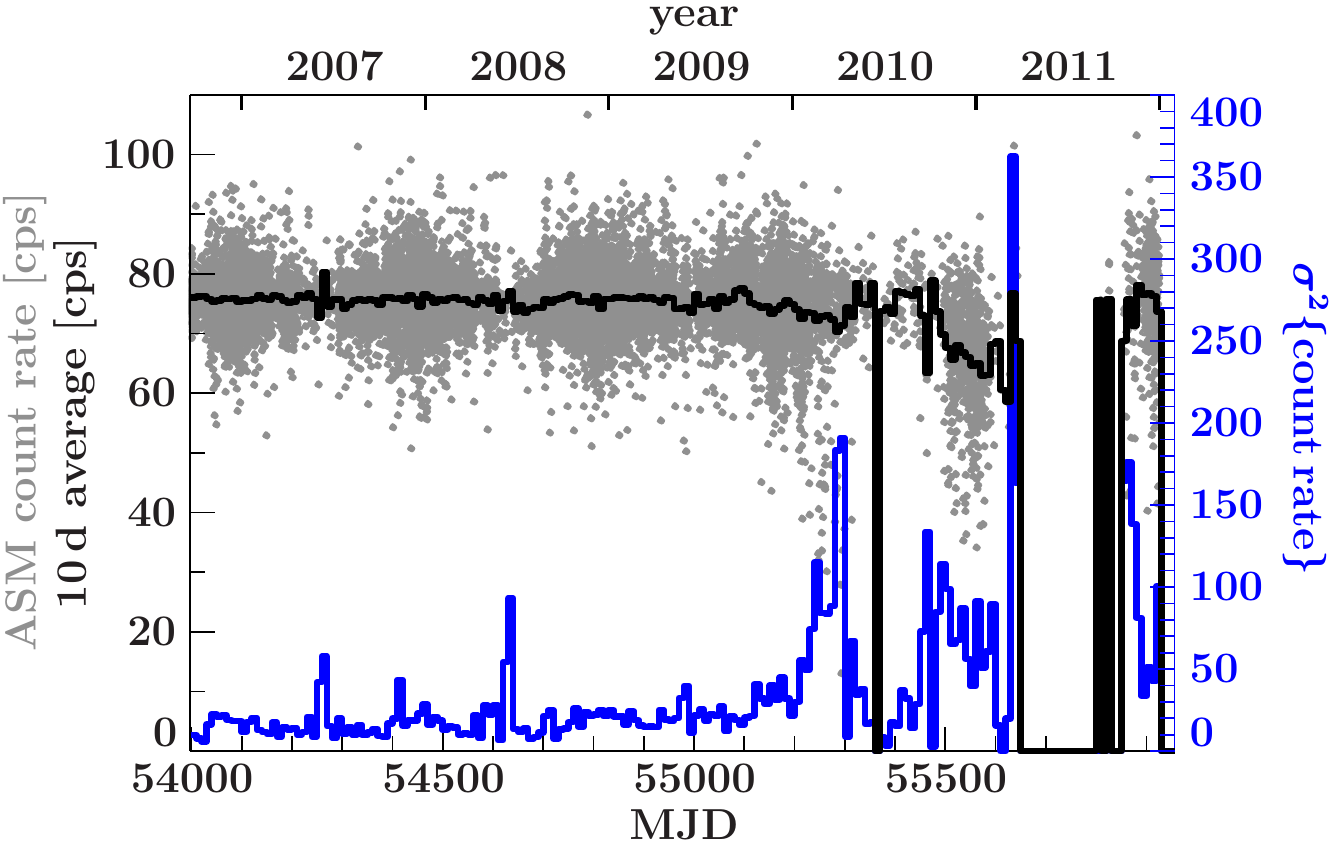}\hfill
\includegraphics[width=\columnwidth]{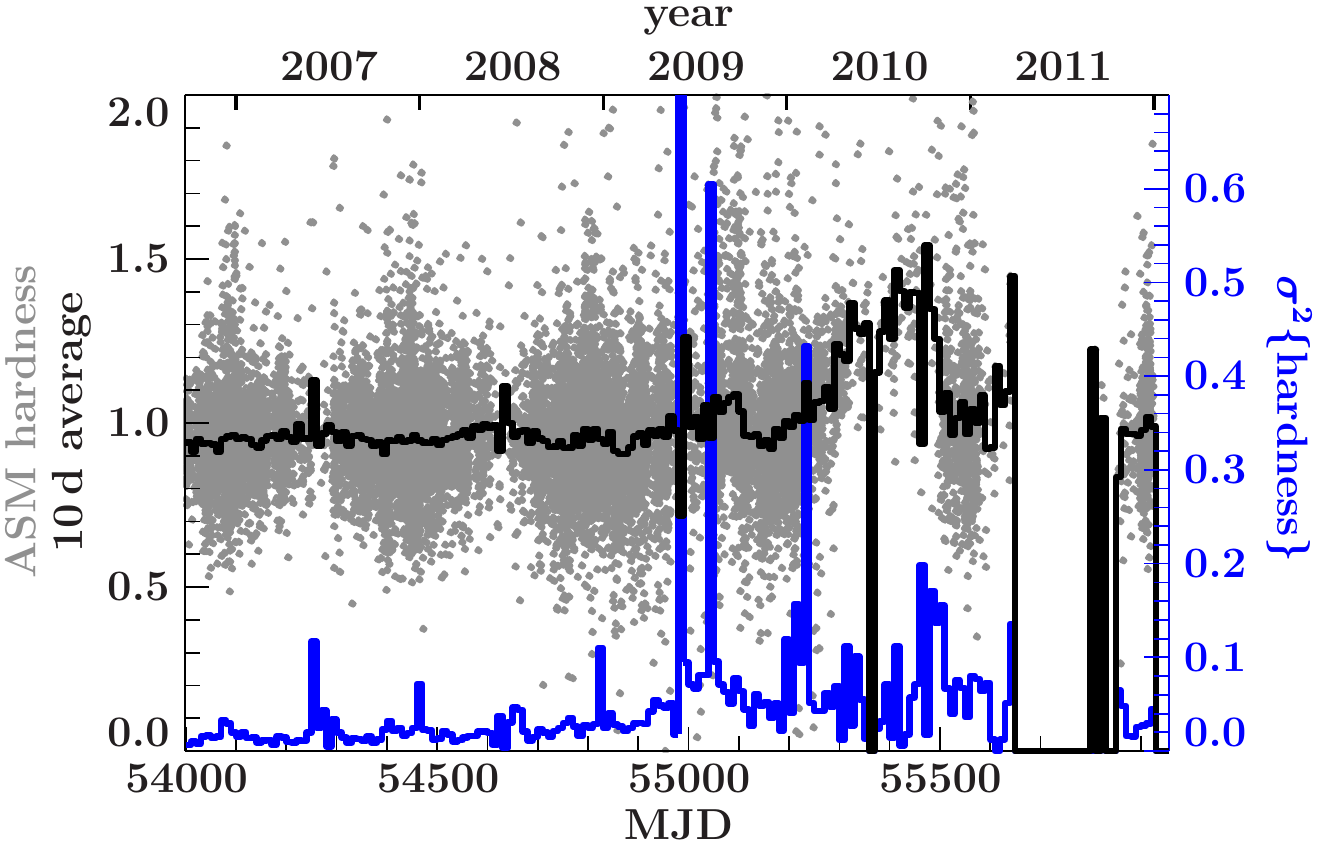}
\caption{Crab ASM light curve (left panel, 1.5--12\,keV, gray points)
  and hardness (right panel, 5.0--12\,keV/1.5--3.0\,keV, gray points)
  with the respective 10\,day average values (black histograms) and
  variances (blue histograms).}\label{fig:crab}
\end{figure*}

We consider all 97556 RXTE All Sky Monitor
\citep[ASM,][]{Levine_1996a} measurements of \mbox{Cyg\,X-1}\xspace
performed during the lifetime of RXTE. In the following all data
analysis was performed with ISIS~1.6.2
\citep{Houck_Denicola_2000a,Houck_2002,Noble_Nowak_2008a}.

We first filter for measurements where the background was clearly
oversubtracted (count rates $r_A$, $r_B$, or $r_C < 0$). The quality
of the ASM data started deteriorating after about MJD~55200 (early
2010~January): valid pointings are fewer and the overall
variance of the measurements becomes larger
(Fig.~\ref{fig:monster}). Since the larger variance could be source
intrinsic, we analyze the ASM light curve of a source known to be
roughly constant at the level of precision required here, the Crab
\citep{Wilson-Hodge_2011a}: after MJD~55200 the Crab light curve shows
prolonged gaps (Fig.~\ref{fig:crab}). Where data exist, the average
values of the ASM count rate (calculated on a 10\,d timescale), which
were stable at $\sim$75\,cps before, decrease by up to 10\%. The
variance of the data strongly increases by a factor of $\sim$10. The
ASM hardness (between the C and~A bands) increases from stable values
around $\sim$0.95 to up to $\sim$1.4 and shows higher variability. The
timescale of this change in behavior as well as the similar behavior
of the light curves of other sources during the same time period imply
that we do not see the long term variability of the Crab nebula as has
been observed by \citet{Wilson-Hodge_2011a} here, but truly
instrumental decline during the last years of the ASM
lifetime.\footnote{\citet{Vrtilek_2012a} come to similar conclusions
  regarding ASM data from 2010 on; they also report on gain changes in
  the last two years of ASM lifetime as a possible cause.}

The ASM data measured post 2010 January can still be used to assess
trends with large changes in count rate; for the following analysis,
however, which relies on absolute values of both count rate and
hardness, we ignore all ASM data after MJD~55200.  Overall, we use
94068 individual ASM measurements corresponding to 80082 individual
times (a source can be in the field of view of two SSCs
simultaneously) spanning over 5000 days from MJD~50087 (1996
January~5) to MJD~55200 (2010 January~4).

\subsection{Pointed RXTE observations}\label{sec:rxte_data}

We consider all pointed RXTE observations of \mbox{Cyg\,X-1}\xspace
made during the RXTE lifetime (MJD~50071 to MJD~55931). Data from the
Proportional Counter Array \citep[PCA,][]{Jahoda_2006a} and the
High-Energy X-Ray Timing Experiment \citep[HEXTE,][]{Rothschild_1998a}
onboard RXTE were reduced with HEASOFT~6.11 following the procedure
introduced in the previous papers of this series
\citep{Pottschmidt_2003b,Wilms_2006a}.

For spectral analysis of the PCA data, we use data from the top xenon
layer of Proportional Counter Unit (PCU)~2 only. PCU~2 is the best
calibrated PCU and had been running during all RXTE observations of
\mbox{Cyg\,X-1}\xspace.  We make use of the improved PCA background
models and therefore only discard data within 10\,minutes of the South
Atlantic Anomaly \citep[SAA;][]{Fuerst_2009a} passages, as opposed to
the 30\,minutes we used previously.

No HEXTE spectra are available during some observations either due
to the non-standard observation mode employed, mainly in early
observations which are not part of our bi-weekly campaign, or due to
the failure of the rocking mechanism of the HEXTE clusters in the last
years of the RXTE lifetime.

We extract all PCA spectra in the \texttt{standard2f} mode for each
RXTE orbit and obtain 2741~individual spectra. The improved PCA
background and response allow us to consider a wider energy range than
previously, namely $\sim$2.8\,keV to 50\,keV. The wider overlap
between the PCA and HEXTE spectra, which are considered between
18\,keV and 250\,keV, leads to a better constraint on the
multiplicative constant which accounts for the differences in the flux
calibration of the two instruments. Following \citet{Boeck_2011a} and
\citet{Hanke_2011_PhD} we add a systematic error of 1\% to the fourth
PCA bin (2.8--3.2\,keV) and of 0.5\% to the fifth PCA bin
(3.2--3.6\,keV) in the data from PCA's calibration epoch 5 (2000
May~13 -- 2012 January~5)\footnote{Over its lifetime the PCA saw four
  different gain calibration epochs, followed by a long fifth epoch
  defined by the loss of the propane layer in PCU0 in 2000 May. See
  \url{http://heasarc.gsfc.nasa.gov/docs/xte/e-c_table.html} for
  details. These epochs are not to be confused with the five activity
  periods of \mbox{Cyg\,X-1}\xspace that we define in
  Fig.~\ref{fig:monster} and Sect.~\ref{sec:general_behaviour}.},
which constitute the majority of our dataset. For epochs~1 to~4 we aim
for the closest possible match in energy for the systematic errors.

Where possible due to the available PCA modes, light curves with a
time resolution of $2^{-9}$\,s ($\sim$2\,ms) were extracted for timing
analysis, choosing the same energy bands as \citet{Boeck_2011a}: a low
energy band corresponding to energies 4.5--5.7\,keV (channels 11--13)
and a high energy band corresponding to energies 9.4--14.8\,keV
(channels 23--35; energy conversion for epoch~5). The calculation of
the rms and X-ray time lags follows \citet{Nowak_1999a}.  We use
segments of 4096 bins, i.e., we calculate the root mean square
variability (rms) between 0.125 and 256\,Hz. 

For two simultaneous, correlated light curves, such as the high and
low energy lightcurves of \mbox{Cyg\,X-1}\xspace used here, one can
calculate a Fourier-frequency dependent time lag between the two from
the Fourier phase lag at any given frequency \citep[][and references
therein]{Nowak_1999a}. In our calculations a positive time lag means
that the lightcurve in the hard energy band is lagging the soft. The
timelag depends strongly on Fourier-frequency and shows a complex
variation with state \citep[see][for examples and note that we use the
same frequency binning in the relevant
frequency-range]{Pottschmidt_2000a}. To obtain a single value which
serves as a good signature for the overall level of the lags, we
average it over the frequency range of 3.2--10\,Hz
\citep[see][]{Pottschmidt_2000a,Pottschmidt_2003b}.

\subsection{Swift-BAT, MAXI, and Fermi-GBM data}\label{sec:bat_maxi}

RXTE was switched off on 2012 January~5 (MJD~~55931). To continue the
monitoring of the long term behavior of \mbox{Cyg\,X-1}\xspace we
therefore need to use other instruments. The all sky monitor available
in the soft X-ray band at the time of writing is MAXI, the hard X-rays
above 10\,keV are covered by Swift-BAT and the Fermi-GBM.

MAXI is an all sky monitor onboard the Japanese module of the
International Space Station \citep{Matsuoka_2009a}. Light Curves from
the Gas Slit Camera detector (GSC) are available in three energy bands
(2--4\,keV, 4--10\,keV and 10--20\,keV) on a dedicated
website\footnote{\url{http://maxi.riken.jp/top/index.php?cid=1&jname=J1958+352}}. MAXI
light curves show prolonged gaps of several days due to observational
constraints.

Swift-BAT is sensitive in the 15--150\,keV regime
\citep{Barthelmy_2005a}. Regularly updated, satellite-orbit averaged
light curves in the 15--50\,keV energy band from this coded mask
instrument are available on a dedicated website
\footnote{\url{http://swift.gsfc.nasa.gov/docs/swift/results/transients/}}.

The Gamma-ray Burst Monitor \citep[GBM;][]{vonKienlin2004,Meegan2007}
onboard Fermi observes the sky in the hard X-ray and soft $\gamma$-ray
regime (about 8\,keV to $\sim$30\,MeV). It permanently provides
complete coverage of the unocculted sky. Due to its strongly limited
spatial resolution, the brightness of individual sources cannot be
determined directly and the Earth occultation method is applied
\citep[][]{Case2011,Wilson-Hodge2012}. In this work we use the
publicly available quick look Fermi GBM Earth occultation results
provided by the Fermi GBM Earth occultation Guest Investigation teams
at NASA/MSFC and LSU\footnote{\url{http://heastro.phys.lsu.edu/gbm/}}
which consist of light curves with a 1\,d resolution in four energy
bands between 12\,keV and 300\,keV (12--25\,keV, 25--50\,keV,
50-100\,keV, and 100--300\,keV) starting from MJD 54690. On average
each measurement of \mbox{Cyg\,X-1}\xspace is based on 18
occultations.

As these data are pre-screened by the respective instrument teams, no
further selection criteria were applied and we use all data available
from the start of the each mission until MJD~56240, resulting in 36454
individual measurements for BAT, 9794 for MAXI, and 1443 for GBM
(Fig.~\ref{fig:monster}).

\section{Identifying the states of Cygnus~X-1}\label{sec:id_states}

\subsection{General source behavior}\label{sec:general_behaviour}

The different periods of source activity and therefore the different
population of the individual source states strongly affect our ability
to distinguish the regions the respective states occupy on
hardness-intensity diagrams (HIDs) or in other spaces. As a first step
for our analysis we therefore present an overview of the light curves
used for this work and over the ASM 5--12\,keV to 1.5--3\,keV hardness
(Fig.~\ref{fig:monster}).

The source behavior from early 1996 (RXTE launch) until the end of
2004 has been discussed by \citet{Wilms_2006a}, who used a crude
definition of ASM based states using the count rate only. This
classification is sufficient to distinguish main activity patterns,
and also consistent with more detailed studies
\citep{Zdziarski_2011a}. We extend this earlier work and supplement
the ASM data with BAT, MAXI, and GBM. By eye, i.e., without using the
state definitions introduced later in this section and shown in color
on Fig.~\ref{fig:monster}, we are able to distinguish five main
periods with different source activity patterns:
\begin{itemize}
\item a pronounced soft state episode in 1996 (up to $\sim$MJD~50350,
  period~\textsc{i}),
\item two and a half years of a mainly stable hard state with only
  short softenings, seen as spiky features on Fig.~\ref{fig:monster},
  between the end of 1996 and early 1998 ($\sim$MJD~50350--51000,
  period~\textsc{ii}),
\item a series of failed state transitions and soft states which
  started in early 1998 \citep{Pottschmidt_2003b} and included a
  prolonged, very soft period between the end 2001 and the end of
  2002. This activity period continued until mid-2006
  ($\sim$MJD~51000--53900, period~\textsc{iii}),
\item an almost continuous hard state from mid-2006 to mid-2010 which
  includes the hardest spectral states ever observed in
  \mbox{Cyg\,X-1}\xspace \citep[see][$\sim$MJD~53900--55375,
  period~\textsc{iv}]{Nowak_2011a},
\item a series of prolonged soft states which followed an abrupt state
  transition at $\sim$MJD~55375 and continued during the writing of
  this paper ($>$MJD~55375, period~\textsc{v}).
\end{itemize}
Periods~\textsc{i} to~\textsc{iii} are fully and period~\textsc{iv} is
almost fully covered by the ASM. Some ASM data exist during
period~\textsc{v}, but are affected by problems described in
Sect.~\ref{sec:asm_data} and therefore excluded from our analysis.

During period~\textsc{iv} the ASM hardness shows higher variability
than in previous hard states, but the same effect is also seen in the
hardness of the Crab and is instrumental (Fig.~\ref{fig:crab}).

BAT coverage started at the end of period~\textsc{iii} on MJD~53414
(mid-February 2005) and continues during the writing of this paper.
Thus, BAT data are available during soft states, namely in period~\textsc{v},
but simultaneous coverage of soft states by BAT and ASM is lacking.

GBM and MAXI coverage started MJD~54690 (2008 August~12) and MJD~55058
(2009 August~15), respectively, and was ongoing as of time of writing.
These instruments mainly cover the final phase of period~\textsc{iv} and the
ongoing period~\textsc{v}. No simultaneous coverage with ASM exists during the
intermediate and soft states, except during the phase of ASM
deterioration.

During the joint coverage by ASM and BAT, \mbox{Cyg\,X-1}\xspace
displayed two softening episodes, which did not reach a stable soft
state ($\sim$MJD~53800--53900 and around MJD~55000). Both episodes are
clearly visible in the ASM band but not in the BAT light curve. This
is a remarkable contrast to the full state transition of
$\sim$MJD~55375, which is associated with a clear drop in the BAT
count rate simultaneous with an increase of ASM count rate. Similar
softenings visible in the soft X-ray band but not in the hard band
have been observed in 1997--1999 by \citet{Zdziarski_2002a}, who
discuss changes seen in the ASM where no corresponding changes were
observed in the 20--300\,keV band with BATSE. Only further long term
observations in both soft and hard X-rays can help to decide whether
only successful state transitions are accompanied by a change in hard
X-ray flux and whether the behavior of the hard component above
15--20\,keV can be used to decide whether a state transition will fail
or not.

After MJD~55200 reliable ASM data are lacking. The BAT light curve
suggests that while the source left the soft state for short periods
of time, those hard periods were softer than the prolonged hard states
in periods~\textsc{ii} and~\textsc{iv}.

A striking feature of the hardness during the hard states are values
which exceed the average by a factor of $\sim$5 and more
\citep{Balucinska_Church_2000a,Poutanen_2008a}. They correspond to
so-called X-ray dips, where blobs of cold material in the line of
sight cover the source. Since it is virtually impossible to decide
whether an individual ASM dwell was measured during a dip, we do not
treat these data separately for the purpose of this paper \citep[but
see][]{boroson_2010a}.

\subsection{$\Gamma_1$ defined states}\label{sec:pca_analysis}

\citet{Wilms_2006a} have shown that the spectral shape of
\mbox{Cyg\,X-1}\xspace as observed by PCA and HEXTE (from here on:
RXTE spectrum) can be well described by an empirical model which
consists of a broken power law modified by a high energy cutoff,
absorption described by the \texttt{tbnew} model, see, e.g.,
\citet{hanke_2009a}, and the abundances of \citet{Wilms_2000a}. The
iron K$\alpha$ line is described by an additive Gaussian emission line
at $\sim$6.4\,keV. The intermediate and soft states additionally show
a soft excess, which is usually described by accretion disk emission
\citep{Mitsuda_1984a,Makishima_1986}. The soft photon index,
$\Gamma_1$, of the broken power law shows strong correlations with
other spectral and timing parameters on timescales from hours to years
across the whole range of its values and is a good proxy for the
source state
\citep[e.g.,][]{Pottschmidt_2003b,Wilms_2006a,Boeck_2011a,Nowak_2011a}.

We model all spectra both with and without a disk and accept the disk
as real if the addition of the disk component improves the $\chi^2$
value by more than 5\%\footnote{Since the PCA data in the lower
  channels are dominated by systematic errors for a source as bright
  as Cyg~X-1, it is not possible to adopt a significance-based
  criterion for the improvement in $\chi^2$.}. With this approach, we
are able to achieve good fits ($\chi^2_{\mathrm{red}} < 1.2$) for
almost all spectra, with a few outliers not exceeding
$\chi^2_{\mathrm{red}} = 2.5$. Less than 18\% of the spectra with
$\Gamma_1 < 2.0$ require a disk, but 97\% of the fits with $\Gamma_1 >
2.0$ do. This agrees with the known behavior of the disk in black hole
binaries in the different states \citep[e.g.,][]{Belloni_2010a}. For
all best fit models, we obtain $\Gamma_1 > \Gamma_2$ and
$E_{\mathrm{break}} \sim 10$\,keV. For a more detailed discussion of
the fits to then available data and examples of typical spectra we
refer to \citet{Wilms_2006a}.

\begin{figure}
\includegraphics[width=\columnwidth]{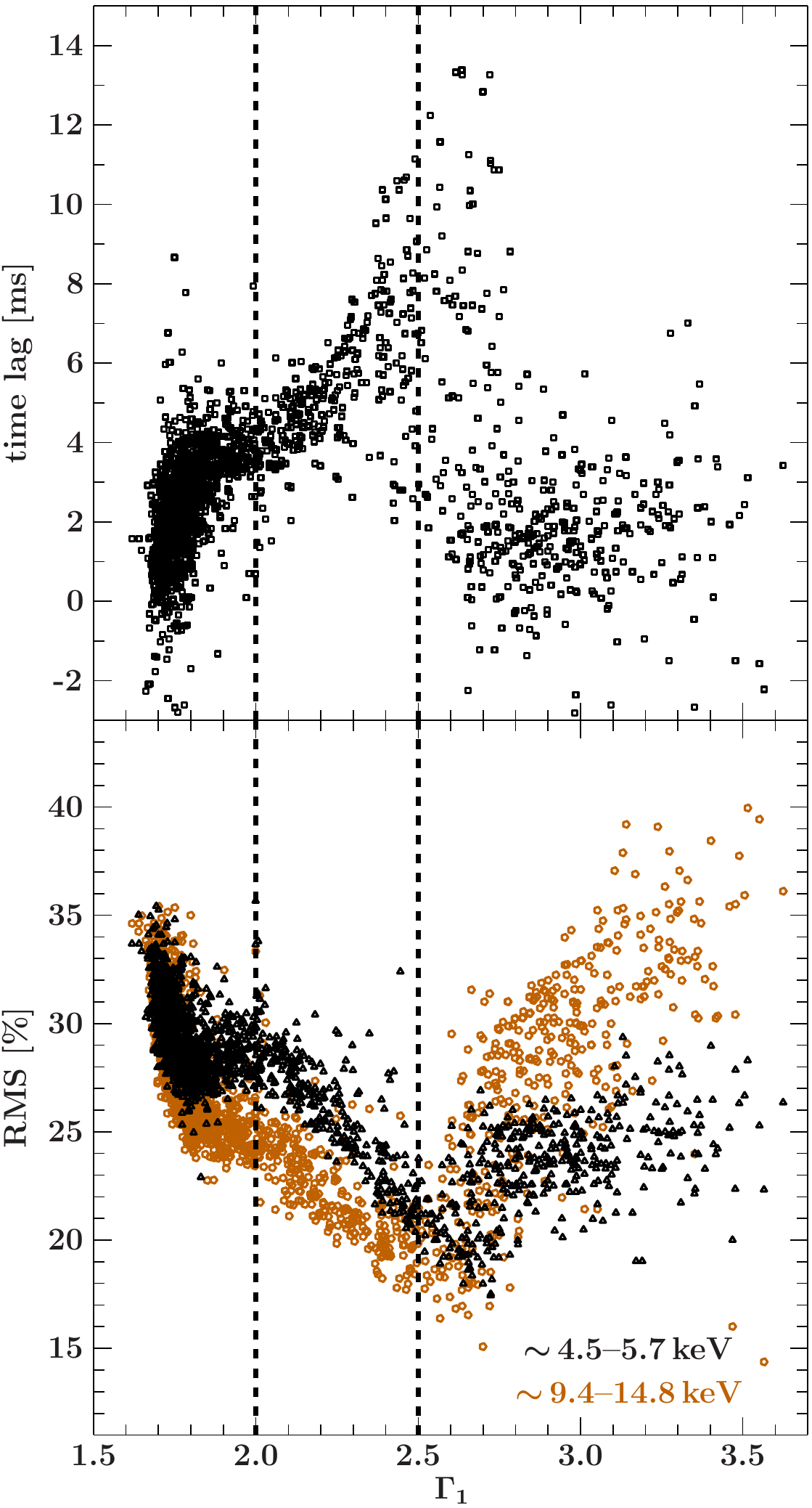}
\caption{X-ray time lag and fractional RMS as a function of the soft
  photon index $\Gamma_1$.}\label{fig:timing}
\end{figure}

Figure~\ref{fig:timing} shows the dependence of the total rms between
0.125 and 256\,Hz in both high and low energy bands and the time lag
(averaged between 3.2 and 10\,Hz) on the soft photon index
$\Gamma_1$. The timing and spectral behavior is clearly related, but
complex with changes in both slope and sign.  For clarity of the plot
we do not show error bars; the uncertainity of single measurements is,
however, on the order of or smaller (in the case of the hard
observations: much smaller) than the spread of the correlation at any
given frequency. $\Gamma_1 \sim 2$, where spectral models with a disk
become dominant, corresponds to a bend in the timing
correlations. Another clear kink can be seen at $\Gamma_1 \sim 2.5$.

We use both our spectral and timing analysis results to
define $\Gamma_1$ ranges for different source states that are
characterized by spectral and timing behavior which is similar within
a state but different between the three states. Hard states
correspond to $\Gamma_1 < 2.0$, intermediate states correspond to $
2.0 < \Gamma_1 < 2.5$, and soft states correspond to $\Gamma_1> 2.5$.

\subsection{Simultaneous ASM mapping}\label{sec:asm_sim}

\begin{figure}
\includegraphics[width=\columnwidth]{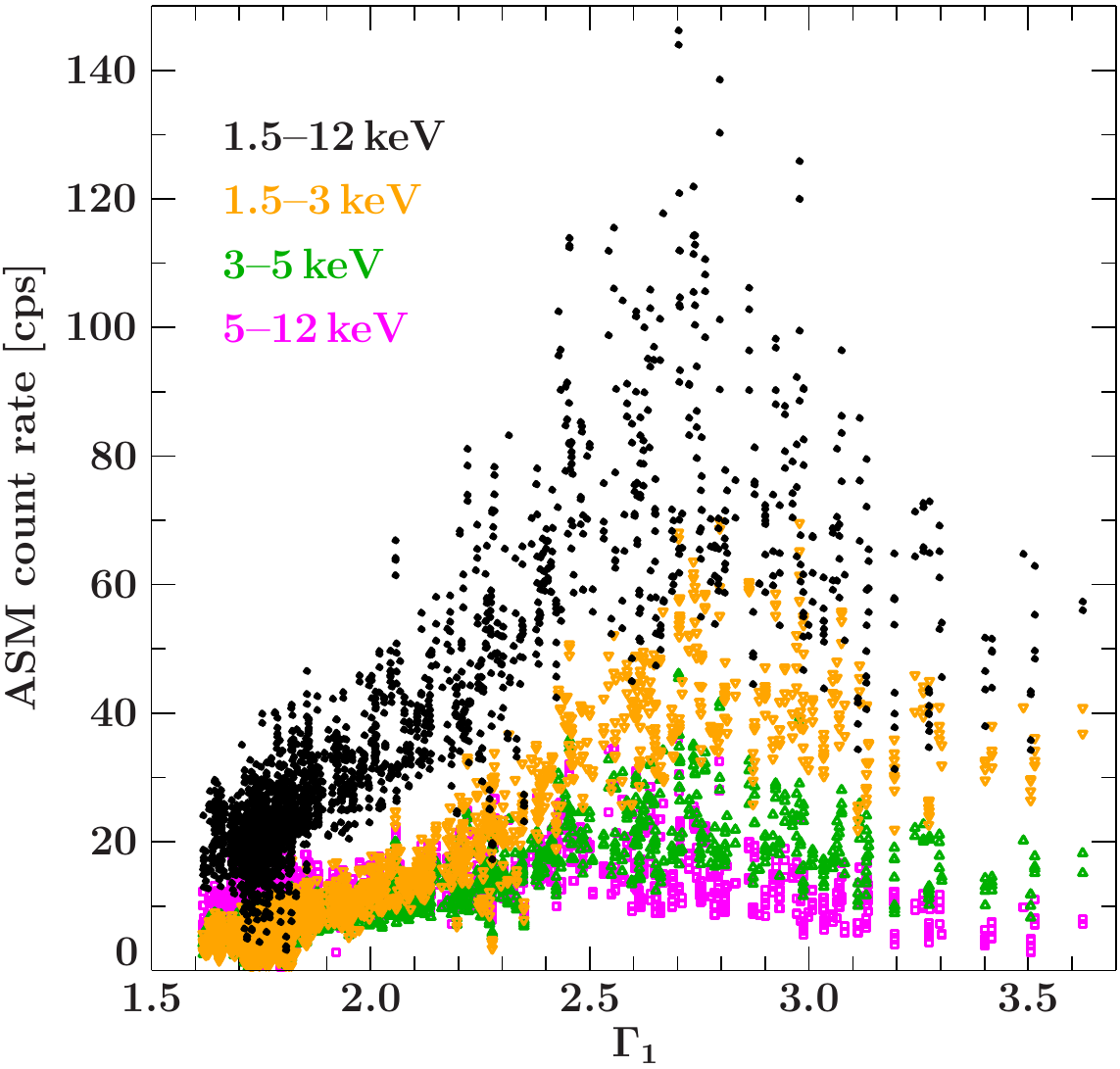}
\caption{Dependence of the ASM count rate in different bands on the
  soft photon index $\Gamma_1$ of simultaneous pointed RXTE
  observations. Shown are total ASM counts (1.5--12\,keV, black dots)
  and counts in the three ASM energy bands: band~A (1.5--3\,keV,
  orange upside down triangles), band~B (3--5\,keV, green triangles)
  and band~C (5--12\,keV, magenta squares).}\label{fig:gamma_asm}
\end{figure}

For direct classification of the ASM data we use the 1424 instances
when ASM observations of \mbox{Cyg\,X-1}\xspace are simultaneous with
spectra from pointed RXTE observations, i.e., fall within the good
time intervals used for the orbit-wise spectral extractions. This
corresponds to 2400 individual ASM measurements since the source is
often observed by more than one SSC. We treat the measurements of
different SSCs independently for the actual mapping since instrument
alignment onboard RXTE would otherwise introduce a bias towards a
higher number of ASM measurements with \mbox{Cyg\,X-1}\xspace in the
field of view of two SSCs during pointed observations of
\mbox{Cyg\,X-1}\xspace.

\begin{table}
  \caption{Spearman's rank correlation coefficients $\rho$ for the
    correlations between the ASM counts and the soft photon index
    $\Gamma_1$ of fits to pointed RXTE observations for the cases
    $\Gamma_1 < 2.7$ and $\Gamma_1 > 2.7$ in ASM 
    energy bands}\label{tab:gamma_asm} 
\begin{tabular}{lcccc}
\hline  \hline
Energy band [keV]
&1.5--12&1.5--3.0&3.0--5.0& 5.0--12\\
\hline
$\rho$ $(\Gamma_1 < 2.7)$& 0.77 & 0.81 & 0.76 & 0.62\\
$\rho$ $(\Gamma_1 > 2.7)$ & $-0.60$ & $-0.54$ & $-0.55$ & $-0.68$\\
\hline
\end{tabular}
\end{table}

Figure~\ref{fig:gamma_asm} shows the dependence of the ASM count rate
from individual ASM-SSC dwells on the soft photon index $\Gamma_1$ of
the broken power law fits. Note that due to the use of individual ASM
SSC dwells, an RXTE spectrum and therefore a $\Gamma_1$-value can
correspond to more than one simultaneous ASM measurement. While the
ASM count rate in all bands is correlated with $\Gamma_1$, this
correlation is not unique: it appears to change sign at $\Gamma_1 \sim
2.7$ (Table~\ref{tab:gamma_asm}). Consistent with this observation,
\citet{Zdziarski_2011a} note that on several instances in the soft
state their derived bolometric flux is lower than during the high-flux
hard state\footnote{\citet{Zdziarski_2011a} define the soft state
  according to a power-law photon index deduced from the ASM count
  rates. This approach does not exactly correspond to the approach
  chosen here, however, it is similar enough to allow rough
  comparisons.}. Thus, cuts in ASM count rate alone cannot separate
the states.

\begin{figure}
  \includegraphics[width=\columnwidth]{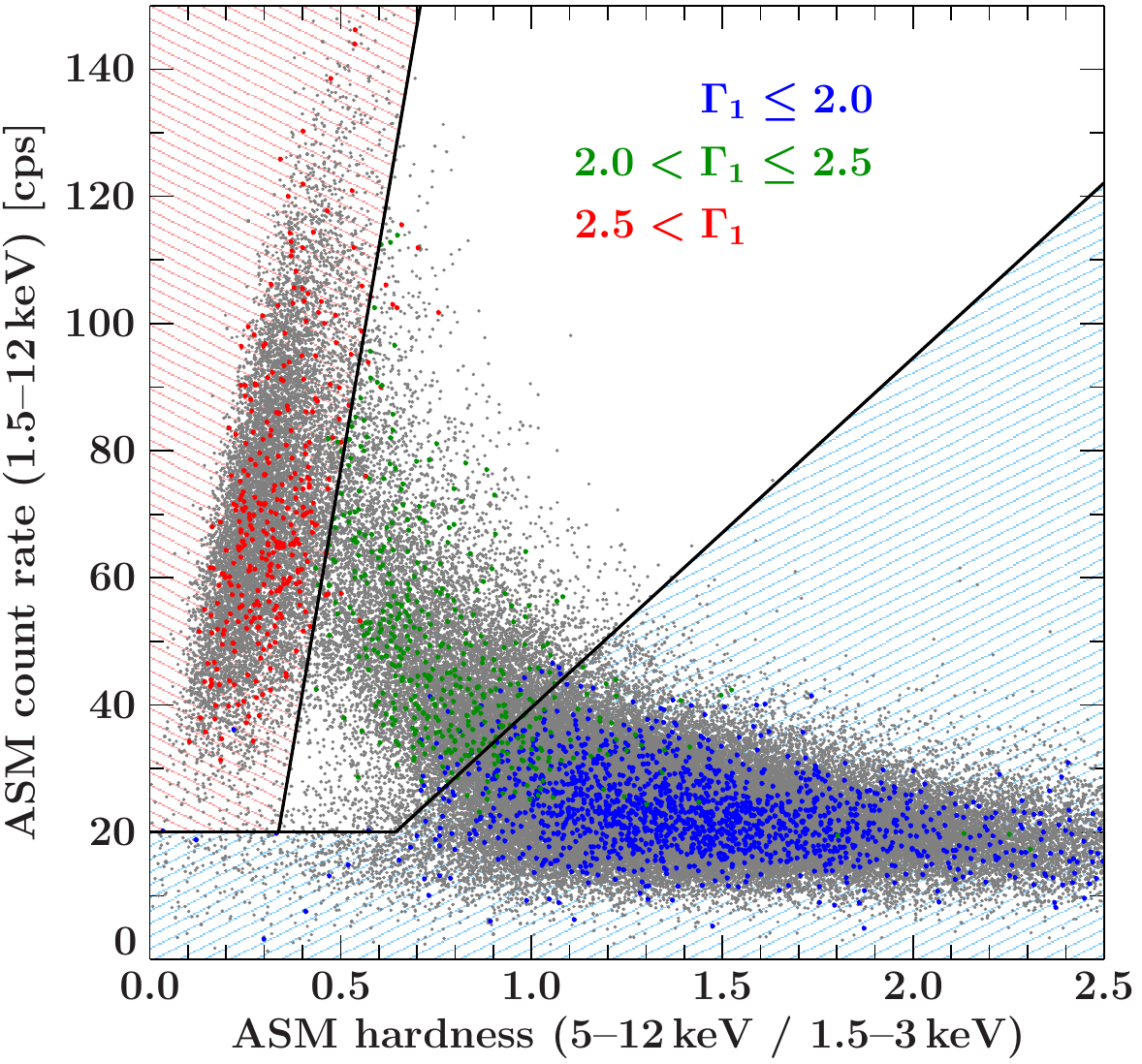}
  \caption{PCA to ASM mapping. Grey data points are all ASM
    measurements of \mbox{Cyg\,X-1}\xspace in the shown range. Blue
    points represent PCA defined hard states, green intermediate and
    red soft states. Black lines show the cuts defining the states in
    the ASM HID. The light blue shaded region corresponds to the
    position of the hard state in the HID, the light red shaded to the
    position of the soft state. The intermediate state region is shown
    without shading.}\label{fig:asm_map}
\end{figure}

The $\Gamma_1$-based definition of states introduced in
Sect.~\ref{sec:pca_analysis} results in in 1608 ASM measurements in
the $\Gamma_1$-defined hard state, in 455 ASM measurements in the
$\Gamma_1$-defined intermediate state, and in 337 ASM measurements in
the $\Gamma_1$-defined soft state. Each of the states primarily
populates a well-defined distinct area in the ASM HID
(Fig.~\ref{fig:asm_map}). A visual inspection reveals that neither
cuts in hardness only nor cuts in count rate only yield a good
division between the states: the softest observations tend to have
count rates in the transitional state range ($\sim$40--50\,cps), the
hard state area extends to low hardnesses usually associated with
transitional or even soft states, and the hard and transitional state
strongly overlap both in ASM hardness and count rate. To account for
these features we choose the following ansatz: an ASM observation with
count rate $c$ and (5--12\,keV/1.5--3\,keV) hardness $h$ is defined as
hard if $c \leqq 20$\,cps regardless of the hardness. For $c >
20$\,cps the cuts between the states are defined by linear functions
of the form
\begin{equation} 
  c_\mathrm{ASM} = m_{\mathrm{ASM}, i} \cdot (h-h_{0,i})
\end{equation} 
where $i\in\{\mathrm{hard},\mathrm{soft}\}$. The line with slope
$m_{\mathrm{ASM}, \mathrm{hard}}$ and $x$-intersection
$h_{0,\mathrm{hard}}$ divides the hard and the intermediate state, and
the line with slope $m_\mathrm{soft}$ and intersection
$h_{0,\mathrm{soft}}$ divides the intermediate and the soft state,
respectively.

We determine the best division between the states such that the
fractional contamination of the ASM-defined states by different
$\Gamma_1$-defined states is minimized. Contamination here is defined
for the hard state (and accordingly for soft and transitional states)
as the fraction of all measurements classified as hard using the ASM
that is classified as transitional or soft according to $\Gamma_1$.
Initial fits indicate that good separations of the states are achieved
for $h_{0,\mathrm{hard}} \sim h_{0,\mathrm{soft}}$. We therefore
reduce the number of free parameters for the cuts and set
\begin{equation} 
  h_{0,\mathrm{hard}} = h_{0,\mathrm{soft}} = h_0.
\end{equation} 
For the best cuts we obtain $h_0 = 0.28$, $m_{\mathrm{hard}} =
55$, and $m_{\mathrm{soft }} = 350$ (Fig.~\ref{fig:asm_map}
and Table~\ref{tap:all_map}).

\begin{table*}
  \caption{Overview over all sky monitor based state definitions for
    \mbox{Cyg\,X-1}\xspace.}\label{tap:all_map}
  \centering
\begin{tabular}{lllll}
   \hline \hline
   State & ASM-based\tablefootmark{a} & MAXI-based\tablefootmark{b} & 
   BAT-based\tablefootmark{c} & GBM-based\tablefootmark{d}\\
   \hline
 hard &  $c \leq 20 \, \vee \, c \leq 55\cdot(h-h_0) $  
 & $c_\mathrm{M} \leq 1.4 \cdot h_\mathrm{M}$ & \ldots \tablefootmark{e} & \ldots  \tablefootmark{f}\\
interm. & $c > 20 \, \land \, 55\cdot(h-h_0) < c  \leq 350\cdot(h-h_0)$ 
 & $ 1.4 \cdot h_\mathrm{M} < c_\mathrm{M} \leq 8/3 \cdot h_\mathrm{M}$ & \ldots \tablefootmark{e} & \ldots \tablefootmark{f} \\
soft & $c > 20 \, \land c > 350\cdot(h-h_0)$ 
 & $ 8/3 \cdot h_\mathrm{M}< c_\mathrm{M}$ & $c_{\mathrm{B}} \leq 
 0.09\,\mathrm{counts}\,\mathrm{cm}^{-2}\,\mathrm{s}^{-1}$ & $f \leq 0.6\,\mathrm{Crab}$\\
 \hline
 \end{tabular}
 \tablefoot{
   \tablefoottext{a}{With ASM 1.5--12\,keV count rate $c$ in $\mathrm{counts}\,\mathrm{s}^{-1}$ 
     , ASM (5--12\,keV/1.5--3\,keV) hardness $h$, and $h_0=0.28$.}
   \tablefoottext{b}{With MAXI 2--4\,keV count rate $c_\mathrm{M}$ in 
     $\mathrm{counts}\,\mathrm{s}^{-1}$  and MAXI (4--10\,keV/2--4\,keV) 
     hardness $h_\mathrm{M}$.}
   \tablefoottext{c}{With BAT normalized 15--50\,keV countrate $c_{\mathrm{B}}$ in $\mathrm{counts}\,\mathrm{cm}^{-2}\,\mathrm{s}^{-1}$.}
   \tablefoottext{d}{With daily GBM 25--50\,keV flux $f$.}
   \tablefoottext{e}{Discriminating between hard and intermediate states 
     is not possible based on BAT lightcurves alone. The source is defined 
     as in BAT-based hard \emph{or} intermediate state for $c_{\mathrm{B}} > 
     0.09\,\mathrm{counts}\,\mathrm{cm}^{-2}\,\mathrm{s}^{-1}$.}
   \tablefoottext{f}{Discriminating between hard and intermediate states 
     is not possible based on GBM daily lightcurves alone. The source is 
     defined as in GBM-based hard \emph{or} intermediate state for 
     $f > 0.6\,\mathrm{Crab}$.}
 }
\end{table*}

The contamination by other $\Gamma_1$-defined states is $<$5\% for the
hard state, $<$10\% for the intermediate state and $<$3\% for the soft
state.  Note that since the source behavior changes continuously from
one state to the other, such that the states do not represent three
distinct, fully independent regimes, we do not expect a perfect
separation of the states \citep[e.g.,][]{Wilms_2006a}. The spread in
count rate and hardness is amplified by the orbital and superorbital
modulations of \mbox{Cyg\,X-1}\xspace and by dips
\citep[e.g.,][]{Poutanen_2008a}. The stronger contamination of the
intermediate state is expected: as a transitional state between the
hard and the soft state, it is short-lived and confined by two
divisional lines. The separation between the hard and the intermediate
state appears especially unclean; we therefore advise to treat the
classification cautiously when an observation is close to this cut.

To test our approach we compare the ASM-based behaviour with the
results of the spectro-timing analysis of the quick, observationally
exceptionally well covered intermediate to soft transition presented
by \citet{Boeck_2011a}. In particular, we can recover the moment of
the transitions at slightly before MJD~53410.

\subsection{Non-simultaneous ASM mapping}\label{sec:nonsimul}

In general, an all sky monitor and a pointed instrument will not
observe a source at the same time, so that we need to assess how well
a given monitor pointing can be used to characterize the source state
during a non-simultaneous observation.

\begin{figure}
\includegraphics[width=\columnwidth]{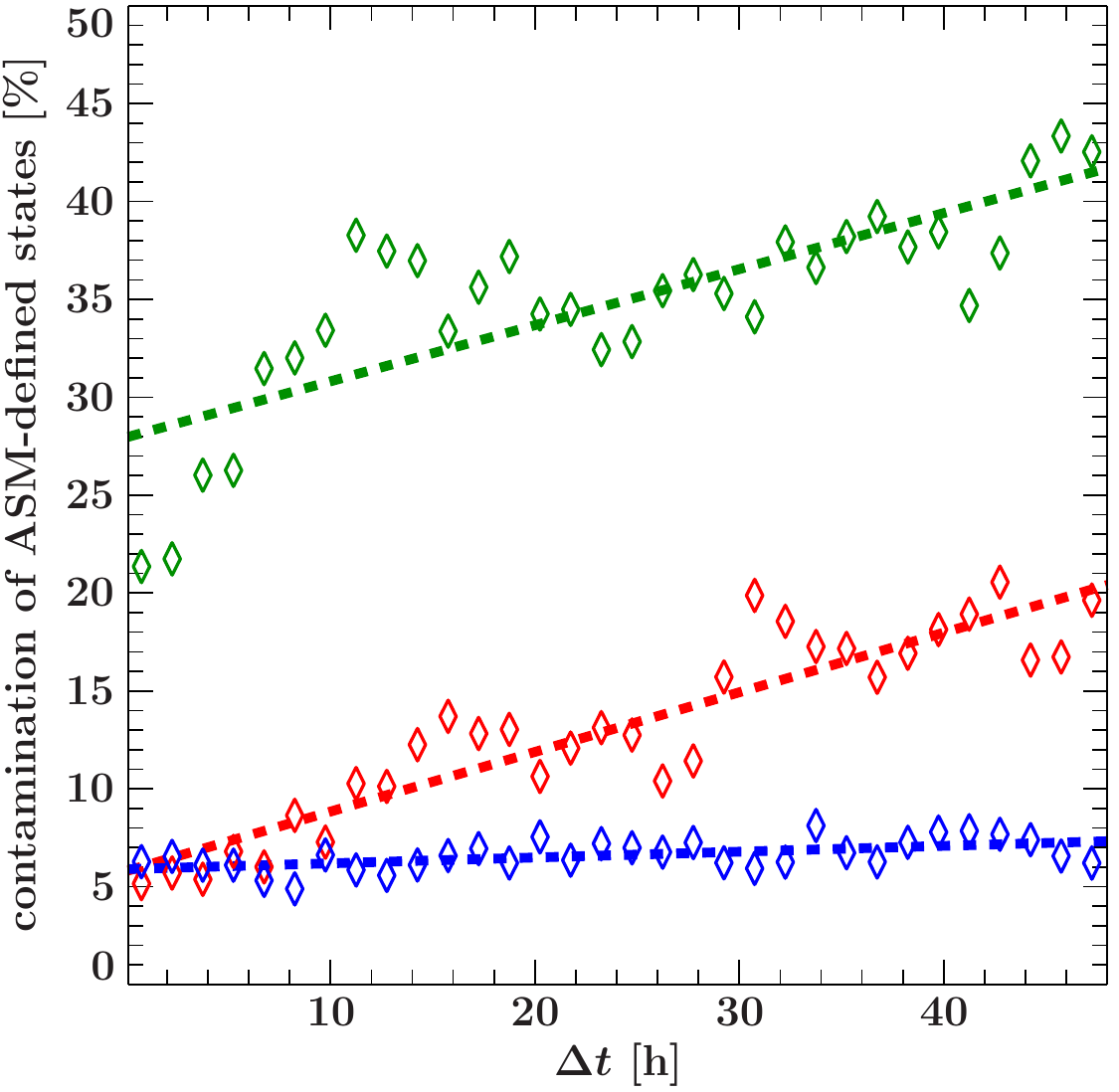}
\caption{Percentage of contamination of the ASM defined states (hard
  state shown in blue, intermediate in green, and soft in red) for
  time delays $\Delta t = \pm (0\text{--}1.5)$\,h,
  $\pm(1.5\text{--}3)$\,h, $\pm(3\text{--}4.5)$\,h, etc., between the
  ASM and the pointed RXTE measurements. Simple linear fits to the
  data are shown as dashed lines to illustrate the overall
  trends.}\label{fig:unsim}
\end{figure}

For every RXTE spectrum we consider ASM measurements within 1.5\,h
intervals $\Delta t = \pm (0\text{--}1.5)$\,h,
$\pm(1.5\text{--}3)$\,h, $\pm(3\text{--}4.5)$\,h, etc., up to
48\,hours. The length of the intervals is motivated by the length of
the RXTE orbit of $\sim$1.5\,h, during about half of which
\mbox{Cyg\,X-1}\xspace is visible. For simultaneous ASM measurements
with different SSCs we use the average for all following analysis. We
obtain 135901 pairs of ASM and RXTE state classifications (the same
ASM measurement may be used to classify several RXTE spectra, if the
RXTE spectra are close enough). For every delay interval and for every
ASM defined state, we determine the percentage of spectra with a
different RXTE classification.

Figure~\ref{fig:unsim} shows that this contamination remains stable
for the hard state for up to $\sim$48\,h. For the soft
state, the contamination reaches 20\% for a 48\,h delay and is even
larger for the intermediate state, which also shows a very strong
increase within the first $\sim$10\,h.  Note that strictly
simultaneous data, which were discussed in Sect.~\ref{sec:asm_sim},
are not taken into account here, resulting in larger starting
contaminations.  The results are similar when using positive or
negative delays only.  The trends are expected, because the hard state
occurred often in long, stable stretches during the RXTE lifetime,
while the intermediate state is short lived due to its transitional
nature.

\subsection{MAXI mapping}\label{sec:maxi}

\begin{figure}
\includegraphics[width=\columnwidth]{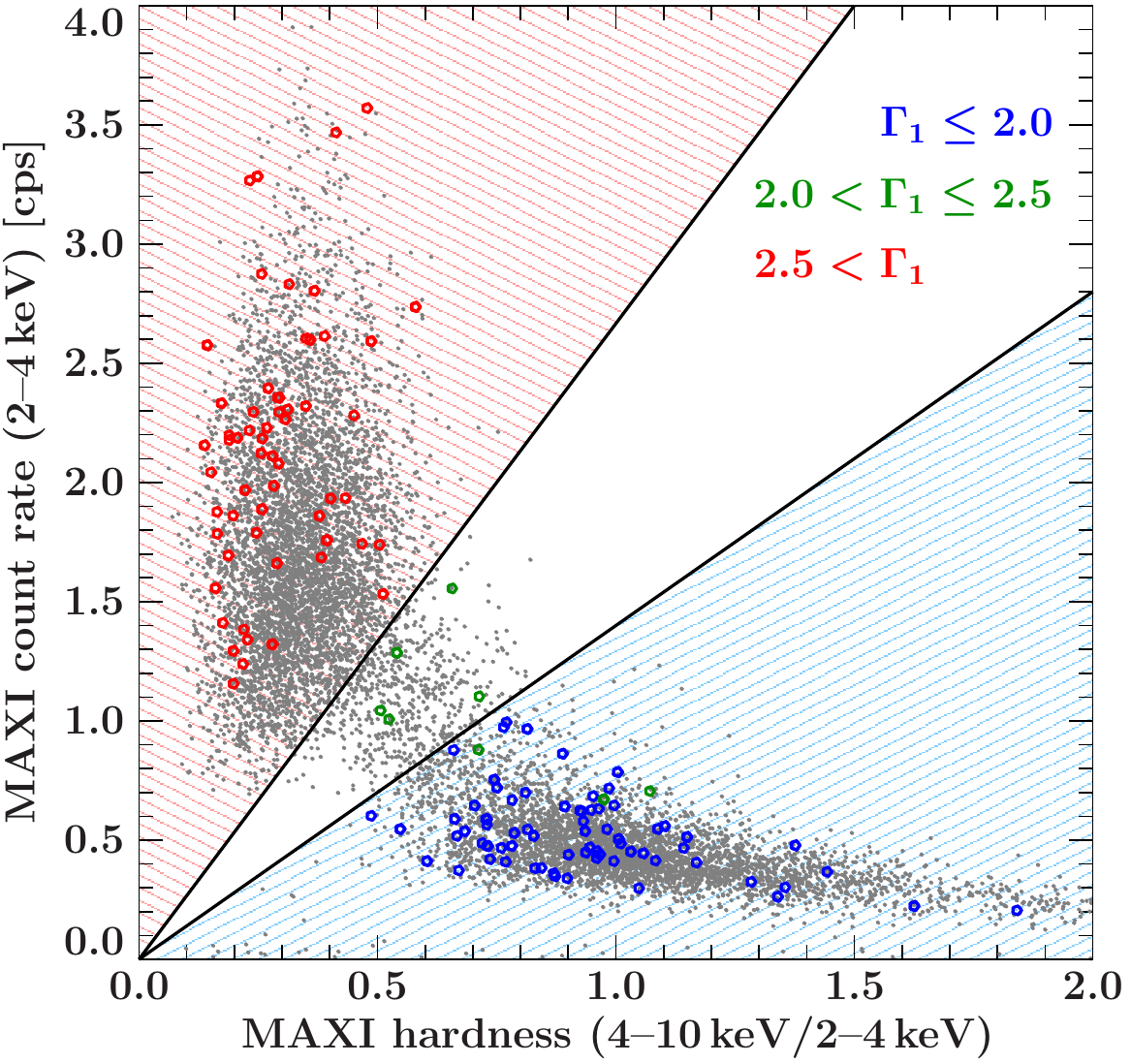}
\caption{PCA to MAXI mapping. Gray data points are all MAXI orbitwise
  measurements of \mbox{Cyg\,X-1}\xspace in the shown energy
  range. Blue circles represent PCA defined hard states, green
  intermediate states and red soft states. Black lines define the
  state cuts. The light blue shaded region corresponds to the hard
  state in the MAXI HID, the light red shaded region to the soft state
  and the region without shading to the intermediate
  state. }\label{fig:maxi}
\end{figure}

Since only 37 MAXI measurements are strictly simultaneous with pointed
RXTE data, we use MAXI data within $\Delta t = 1$\,h before and after
a pointed RXTE observation and obtain 219 MAXI measurements with
$\Gamma_1$-based state classifications, which offer a better overall
statistics with 107~hard states, 101~soft states, and 11~intermediate
states. Increasing $\Delta t$ to 2\,h does not yield a better coverage
of the intermediate state, neither does attempting to map ASM onto
MAXI, since a simultaneous coverage by the two instruments is only
available during the end of period~\textsc{iv}.

To find the best approach to define MAXI based states we consider all
possible different combinations of count rates measured with MAXI (the
three energy bands as introduced in Sect.~\ref{sec:bat_maxi} and the
overall count rate) and different combinations of hardness measures
within the MAXI bands. The clearest separation can be achieved when
using the ratio between count rates in the medium (4--10\,keV) and the
low (2--4\,keV) MAXI bands (MAXI-hardness, $h_\mathrm{MAXI}$) and
the count rate in the low (2--4\,keV) MAXI band, $c_\mathrm{MAXI}$,
as done in Fig.~\ref{fig:maxi}. The MAXI 4--10\,keV/2--4\,keV hardness
is also the closest correspondence to ASM hardness we can achieve
using publicly available MAXI light curves.

The sparse coverage of the intermediate state makes it hard to
separate the three basic states. Given that the hard and the soft
state populate distinct parts of the MAXI HID (see
Fig.~\ref{fig:maxi}) and knowing the shape of the cuts in the ASM HID,
we separate the states by two linear functions of the form
\begin{equation}\label{eq:maxi}
c_\mathrm{MAXI} = m_{\mathrm{MAXI}, i}\cdot h_\mathrm{MAXI},
\end{equation}
where $i\in\{\mathrm{hard},\mathrm{soft}\}$ and where
$m_\mathrm{hard}$ separates the hard and the intermediate state and
$m_\mathrm{soft}$ separates the intermediate and the soft state. The
absence of an $x$-intersection and of a threshold count rate value is
motivated by the lower number of classified data points compared to the
ASM HID, which also prevents us from direct fits for $m$. The best
values obtained by eye are $m_{\mathrm{MAXI}, \mathrm{hard}} = 1.4$ and
$m_{\mathrm{MAXI}, \mathrm{soft}} = 8/3$ (Table~\ref{tap:all_map}).

For a reader interested in performing her own classification using
MAXI data, we note that these are conservative cuts in the sense that
we obtain the purest intermediate state possible here. As in the case
of RXTE-ASM we expect the separation between the hard and intermediate
state to be especially unclean.

\subsection{BAT mapping}

\begin{figure}
\includegraphics[width=\columnwidth]{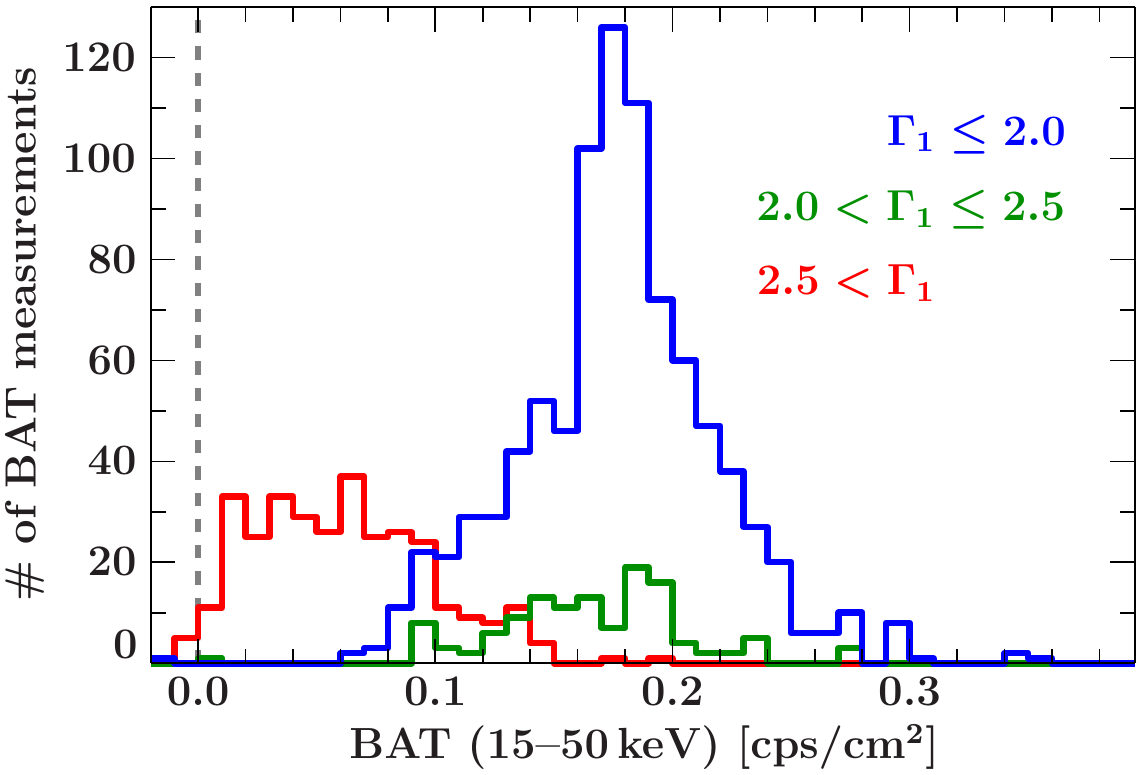}
\caption{Histogram of BAT fluxes of $\Gamma_1$-defined states: the hard state
 is shown in blue, the intermediate state in green, and the soft state in
 red.}\label{fig:bat_hist}
\end{figure}

Swift-BAT light curves in different
energy bands are not readily available. Since BAT also does not cover
soft X-ray energies where the contribution of the accretion disk
becomes important as the spectrum softens, we use only the light curves
in the standard 15--50\,keV band for our analysis and therefore do not
perform a two-dimensional mapping.

Since only 290 BAT measurements are strictly simultaneous with pointed
RXTE data, we employ the same approach as for MAXI
(Sect.~\ref{sec:maxi}) and use BAT data within $\Delta t = 1$\,h
before and after the RXTE spectrum to obtain 1339 BAT measurements
with an RXTE based state classification: 895 in the hard state, 124 in
the intermediate state, and 320 in the soft state.

Only 290 BAT measurements are strictly simultaneous with pointed RXTE
data. Following our study of the reliability of non-simultaneous data
(Sect.~\ref{sec:nonsimul}), we use BAT data within $\Delta t = 1$\,h
before and after a pointed RXTE observation and obtain 1339 BAT
measurements with $\Gamma_1$-based state classifications, which offer
a better overall statistic.

Figure~\ref{fig:bat_hist} shows the histogram of the BAT fluxes in the
different states. While the soft and hard state show clearly different
BAT fluxes, the intermediate states populate the same region as the
hard ones. The BAT measurements, therefore, do not enable us to
separate the hard and the intermediate states. The soft state can
still be identified by a cut at a BAT area normalized count rate of
$c_\mathrm{BAT}=0.09\,\mathrm{counts}\,\mathrm{cm}^{-2}\,\mathrm{s}^{-1}$
For BAT count rates below $c_\mathrm{BAT}$ \mbox{Cyg\,X-1}\xspace is
in the soft state, for fluxes above it in the hard or intermediate
state (Table~\ref{tap:all_map}). The contamination by the other state
is $\sim$7\% in both cases.

We also considered an ASM to BAT mapping where we classified BAT data
by using the closest ASM measurements if one exists within $\pm
0.5$\,h around the BAT measurement in order to increase the number of
data points with classification. No simultaneous good ASM and BAT data
are available in the soft state (see Fig.~\ref{fig:monster} and
Sect.~\ref{sec:asm_data}). The histograms of the hard and intermediate
state follow the trends apparent from Fig.~\ref{fig:bat_hist}: it is
not possible to separate the two states using BAT flux.

\subsection{GBM mapping}

Since GBM lightcurves are publicly available in daily bins only,
strictly simultaneous mapping between RXTE spectra and GBM is not
possible. We therefore define an RXTE spectrum as simultaneous to a
GBM measurement if the good time interval used to extract the spectrum
lies within the GBM measurement. If all simultaneous RXTE spectra show
the same state, the GBM measurement is classified as belonging to this
state. If the states of the RXTE spectra differ, the GBM measurement
is not classified.  

The 104 GBM measurements with simultaneous RXTE observations include
on average 3.8 (reaching from 1 to 8) individual RXTE spectra. Only 6
out of the 104 measurements are unclassified due to state ambiguities.
The remaining valid classifications include 61 hard states, 3
intermediate states, and 34 soft states. All 6 unclassified
measurements include hard as well as intermediate state
classifications. There were no GBM measurements including both
intermediate and soft, or hard and soft RXTE spectra. The distribution
of these results is consistent with \mbox{Cyg\,X-1}\xspace spending
most of the time covered by GBM observations in stable hard and soft
states. That 6 out of 9 GBM measurements including a
$\Gamma_1$-defined intermediate state also include $\Gamma_1$-defined
hard states reflects the high variability and instability of the
intermediate states.

\begin{figure*}
\includegraphics[width=\columnwidth]{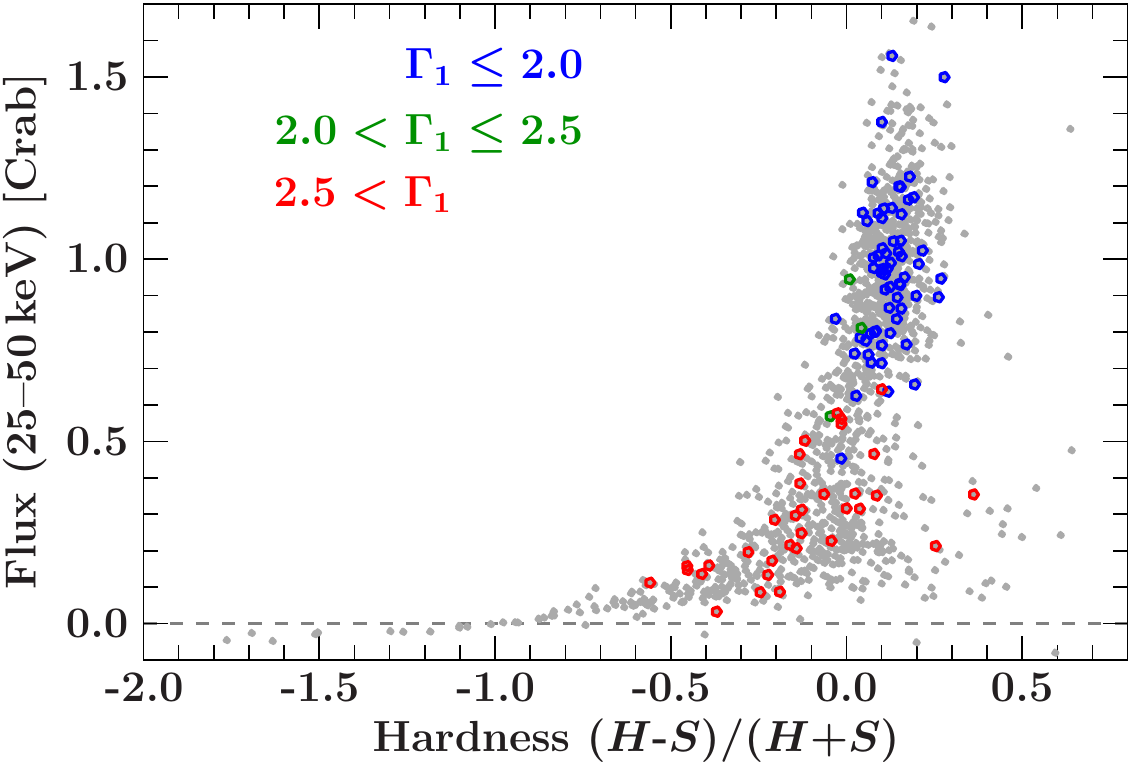}\hfill
\includegraphics[width=\columnwidth]{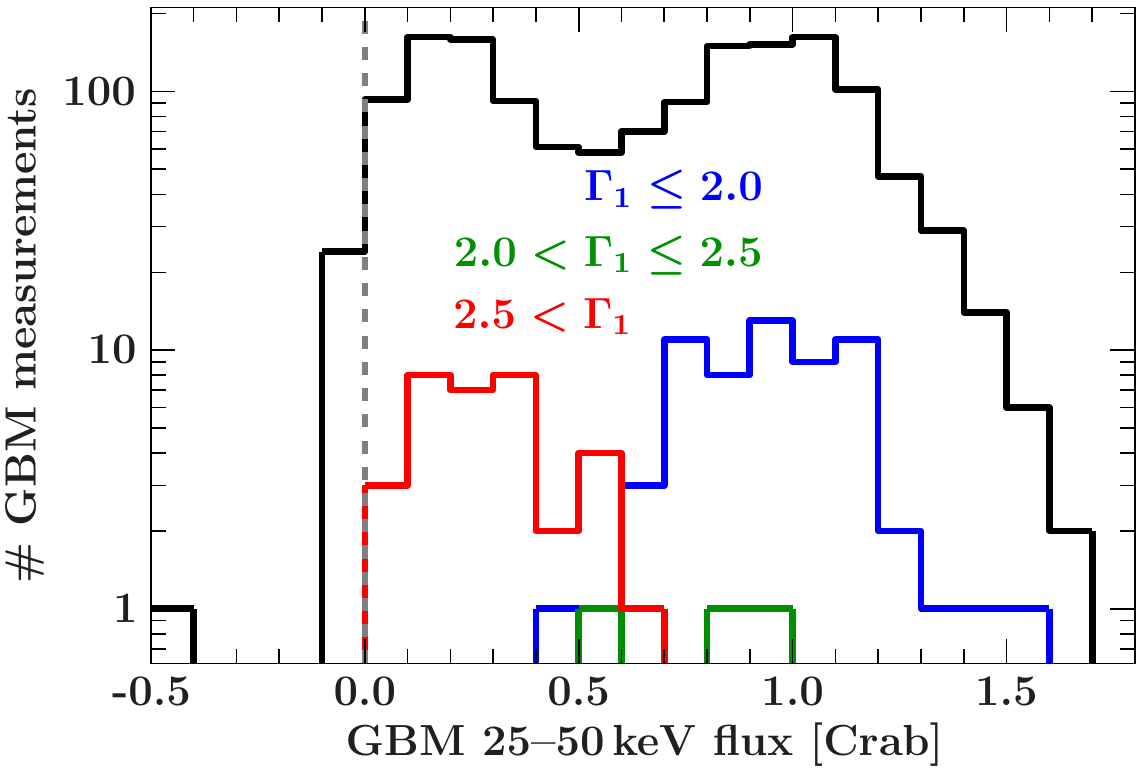}
\caption{\textsl{Left:} Daily GBM HID. $S$ denotes the 12--25\,keV
  flux, $H$ the 25--50\,keV flux. The gray dashed line denotes the
  zero flux level. Total GBM measurements are shown in
  gray. \textsl{Right:} Histogram of daily GBM fluxes of $\Gamma_1$-defined
  states. The hard state is shown in blue, intermediate state in
  green, and soft state in red. Total GBM measurement are shown in
  black. The gray dashed line denotes the zero flux
  level.}\label{fig:gbm}
\end{figure*}

We first consider GBM HIDs. These show a clear separation in two
regions: a region with mainly soft and a region with hard and
intermediate observations, which are clearly separated in GBM flux
only. We therefore consider GBM fluxes only and calculate histograms
of GBM fluxes of $\Gamma_1$-defined states. The best separation
between the states can be achieved when using 25--50\,keV fluxes
(Fig.~\ref{fig:gbm}). While hard and intermediate states cannot be
separated, the soft state can be divided from the other two by a cut
at a GBM flux of 0.6\,Crab: for fluxes $<$0.6\,Crab the source is in
the soft state, for fluxes $\ge$0.6\,Crab in hard or intermediate
state (Table~\ref{tap:all_map}). We do not give values for
contamination here because of the long integration times of the GBM
and because we excluded states where we know the source to change the
$\Gamma_1$ defined state during one GBM measurement. We also note,
that since the hard and intermediate states cannot be distinguished
using GBM, we do not introduce additional bias into our classification
by not including GBM-measurements with simultaneous RXTE spectra in
both hard and intermediate states.

\section{The states of \mbox{Cyg\,X-1}\xspace}\label{sec:states}

\subsection{The statistics of \mbox{Cyg\,X-1}\xspace
  states}\label{sec:statestat}

\begin{table*}
  \caption{Time \mbox{Cyg\,X-1}\xspace spent in the different states (hard -- H,
    intermediate -- I, soft -- S) as measured by the RXTE-ASM,  
    Swift-BAT, MAXI and Fermi-GBM all sky monitors.
    \label{tab:activity_periods}}
\centering
  \begin{tabular}{lccccccccccc}
    \hline \hline
    Period\tablefootmark{a} & MJD & \multicolumn{3}{c}{RXTE-ASM} & 
    \multicolumn{3}{c}{MAXI} &
    \multicolumn{2}{c}{Swift-BAT} & \multicolumn{2}{c}{Fermi-GBM}\\
    && H & I & S  & H & I & S & H \& I & S & H \& I & S\\
    \hline
    I &50087\tablefootmark{b}--50350& 37\% & 19\% & 44\% & \ldots & \ldots & \ldots & \ldots & \ldots & \ldots & \ldots\\ 
    II &50350--51000& 99\% & 1.0\% & 0\% & \ldots & \ldots & \ldots & \ldots & \ldots & \ldots & \ldots\\
    III &51000--53900& 63\% & 20\% & 17\% & \ldots & \ldots & \ldots & \ldots & \ldots & \ldots & \ldots\\
    IV &53900--55375& 97\%\tablefootmark{d}
    & 2\%\tablefootmark{d}
    & $<$1\%\tablefootmark{d} & \ldots & \ldots & \ldots& 99\% & 1\% & \ldots & \ldots \\
    V &55375--56240\tablefootmark{c}& \ldots & \ldots & \ldots &  17\%\tablefootmark{e} & 8\%\tablefootmark{e} & 75\%\tablefootmark{e} & 40\% & 60\% &
    23\%\tablefootmark{f} & 77\%\tablefootmark{f} \\
    \hline
  \end{tabular}
  \tablefoot{
    Between MJD~50087 and 55200 \mbox{Cyg\,X-1}\xspace spent a total of 75.6\% of its 
    time in the hard
    state, 12.8\% in the intermediate state and 11.6\% in the soft state
    as measured with ASM.\\
    Percentages are only shown if at least most of the period is covered by
    the respective instrument.
    \tablefoottext{a}{Periods defined in Sect.~\ref{sec:general_behaviour}.}
    \tablefoottext{b}{Start of period~\textsc{i} defined by the start of the
      RXTE-ASM   measurements.} 
   \tablefoottext{c}{The end of period~\textsc{v} is defined by the
     availability of data at the time of writing.} 
    \tablefoottext{d}{ASM data only until MJD~55200.}
    \tablefoottext{e}{MAXI data affected by gaps in the lightcurve.}
    \tablefoottext{f}{Daily average values are used for GBM.}
 }
\end{table*}

Using the classification from the different instruments derived in the
previous sections, we can assess the activity pattern during the
different periods defined in Sect.~\ref{sec:general_behaviour}. A
detailed break-down of the occurence of different states is given in
Table~\ref{tab:activity_periods}.

Period~\textsc{i} contained a prolonged soft state and the source
spent twice as much time in soft states than in the intermediate state
during this time. We lack all sky monitor coverage before MJD~50087
and therefore do not know whether this soft period observed was a part
of a longer series of soft states. 

In period~\textsc{iii} the source
spent as much time in the intermediate state as in the soft state, in
accordance with the observation of multiple failed state transitions
\citep[e.g.,][]{Pottschmidt_2003b}.

Periods~\textsc{ii} and~\textsc{iv} showed similar activity patterns
dominated by a long, stable hard state, although data from the ASM
were already affected by deterioration of the instrument in
period~\textsc{iv} (Sect.~\ref{sec:general_behaviour}). The state
classifications based on the ASM and the BAT agree well for
period~\textsc{iv}.

The disagreement between the statistics derived from Swift-BAT and
MAXI classification for period~\textsc{v} first seems worrisome. A
visual inspection of Fig.~\ref{fig:monster} indicates, however, that
this mismatch may be due to the gaps in the MAXI light curve. Indeed,
out of the 7794 BAT measurements which fall into MAXI gaps (defined
here as times when the interval between two MAXI measurements is
larger than 6\,h), 3876 are hard or intermediate and 3918 are soft,
while 60\% of all BAT measurements during period~\textsc{v} are soft.
MAXI gaps therefore fall more often onto hard/intermediate states than
onto soft states during this period and the difference seen in
Table~\ref{tab:activity_periods} is due to the incomplete MAXI
coverage. MAXI data can therefore be used to determine states at
certain times, but not to investigate the overall frequency of the
occurrence of different states. A similar caution also applies to the
GBM. Here, the disagreement between the GBM data and other instruments
is partly due to the use of daily lightcurves, where the short term
variability is averaged out (Fig.~\ref{fig:monster}). This can be
directly seen from the higher resolution BAT data. Using individual
BAT points during period~\textsc{v}, 60\% of all measurements are
classified as coming from the soft state
(Table~\ref{tab:activity_periods}). Classifying the daily BAT fluxes,
however, increases the fraction of soft states to 68\%, which is
consistent with the GBM considering the uncertainties in both the GBM-
and BAT-based definitions (Figs.~\ref{fig:bat_hist}
and~\ref{fig:gbm}). Data with a time resolution $<$1\,d are therefore
crucial for a reliable state classification.

Summarizing these observations, we estimate the typical
inter-instrument systematic error of the state determination to be
better than 10\% for the post-RXTE instruments.

\subsection{Stability of States}

\begin{figure}
 \includegraphics[width=\columnwidth]{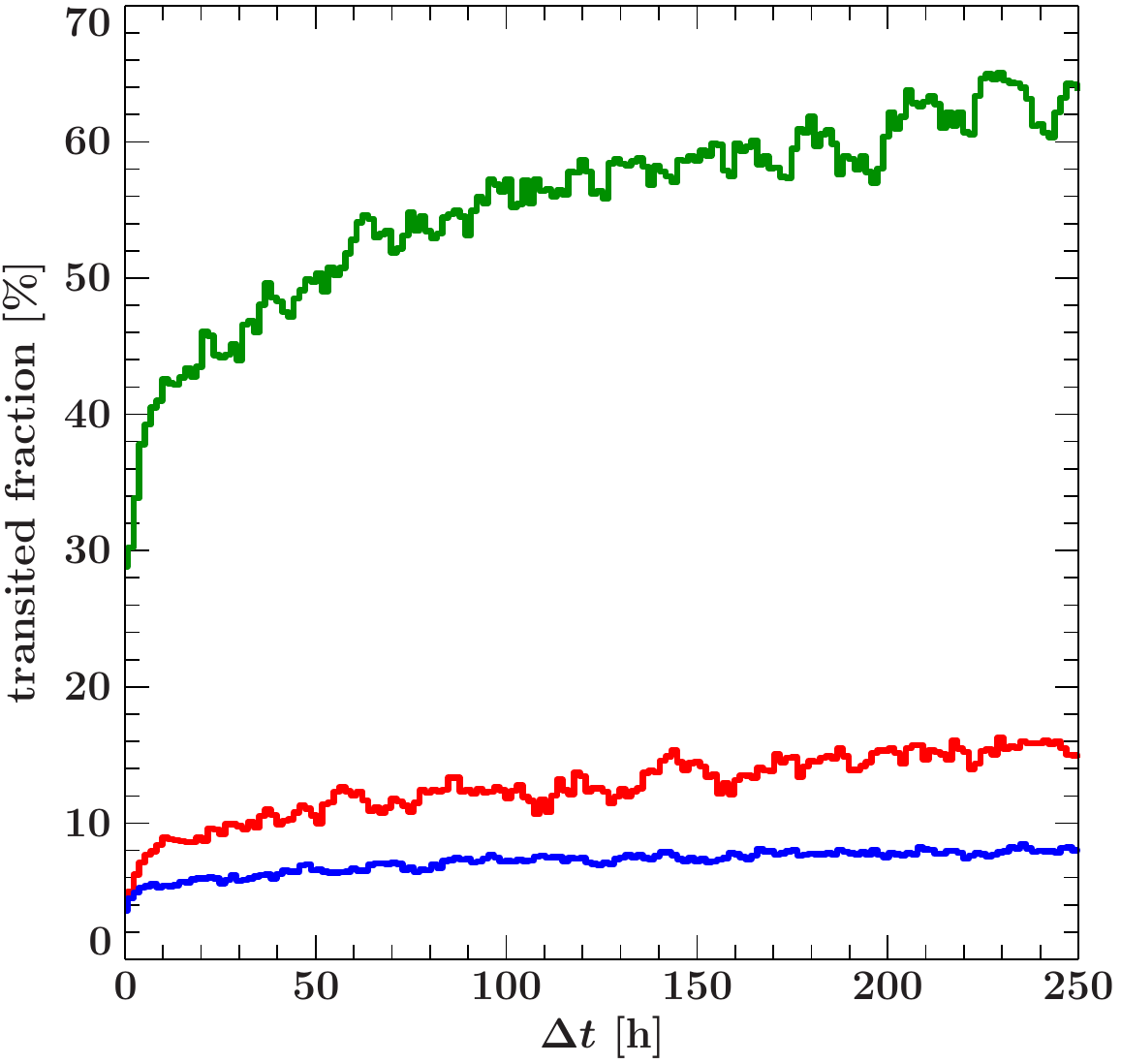}
 \caption{Probability, $P_\mathrm{trans}$, that the source state is
   found changed a time interval $\Delta t$ after a previous state
   determination for the different states (blue: hard, green:
   intermediate, red: soft) using ASM data.}\label{fig:noncumulative}
\end{figure}

\begin{figure}
 \includegraphics[width=\columnwidth]{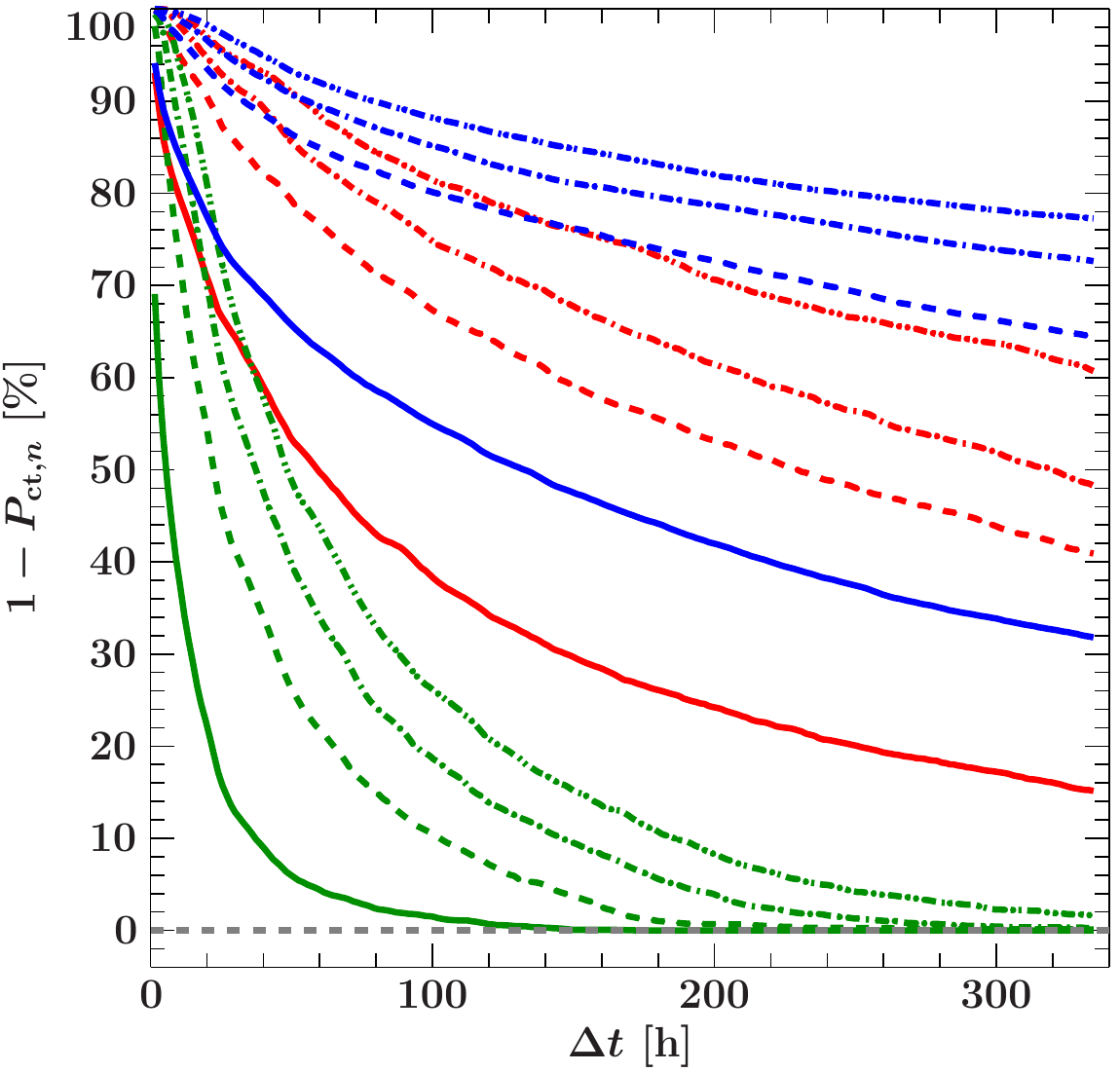}
 \caption{ The probability, $1-P_{\mathrm{ct},n}$, that the source has
   remained in the same state (blue: hard, green: intermediate, red:
   soft) for at least the time $\Delta t$ for different values of $n$
   of ignored number of possible outliers/misclassifications ($n=0$:
   solid, $n=4$: dashed, $n=8$: dot-dashed, $n=12$:
   dot-dot-dashed). The gray dashed line represents 0.
 }\label{fig:cumulative}
\end{figure}

Sect.~\ref{sec:statestat} highlights that
the different states are clearly different in their stability. While
the source tends to remain in the hard and the soft state for a
prolonged time, not unexpectedly the intermediate state is far more
unstable. This well known general behavior can be quantified using the
RXTE-ASM data, the data set in our measurements that covers the
longest time span, does not have gaps, and can be reliably used to
define all three states.

To calculate the state statistics, for each ASM data point measured at
a reference time, $t_\mathrm{ref}$, we calculate the probability that
\mbox{Cyg\,X-1}\xspace has changed its state after a certain interval,
and the probability that the source remained in the same state
throughout this interval. More formally, we calculate the transited
fraction, $P_\mathrm{trans}$, i.e., the probability that an
observation made a certain time \emph{after} the reference observation
will find \mbox{Cyg\,X-1}\xspace in a different source state
(Fig.~\ref{fig:noncumulative}). The probability that the source state
is unchanged is $1-P_\mathrm{trans}$. We determine $P_\mathrm{trans}$
by calculating the fraction of all measurements in a different state
than the reference measurement made at times $t_i$ with $\Delta t_1
\le t_i-t_\mathrm{ref} < \Delta t_2$ where $\{\Delta t_1,\Delta t_2\}
\in \{\{0 \text{\,h},1.5 \text{\,h}\}, \{1.5 \text{\,h},3.0
\text{\,h}\}, \{3.0 \text{\,h},4.5 \text{\,h}\}, \ldots\}$. This
approach is equivalent to that employed in Sect.~\ref{sec:nonsimul},
where we used RXTE spectra as reference measurements.

$P_\mathrm{trans}$ does not take into account the possibility that the
source might have undergone state changes between $t_\mathrm{ref}$ and
$t_i$. To address this possibility we calculate the cumulative
transited fraction, $P_{\mathrm{ct},n}$, which measures the
probability that at least one source transition has happened up to a
time interval $\Delta t$ after the reference measurement. In order to
assess the influence of possible outliers and/or misclassifications,
we define a state transition by the existence of more than $n \in
\{0,\ldots,12\}$ measurements in a different state than the reference
measurement in the time interval $t_\mathrm{ref}\le t <
t_\mathrm{ref}+\Delta t$. The probability that the source has remained
in the same state for at least a time $\Delta t$ is then given by
$1-P_{\mathrm{ct},n}$ (Fig.~\ref{fig:cumulative}).

As expected from a simple look at the lightcurves (see
Fig.~\ref{fig:monster}) hard states are the most stable states: in
more than 90\% of all cases, 200\,h after a given hard state ASM
measurement the source is still in the hard state
(Fig.~\ref{fig:noncumulative}). In over 30\% of all cases a period of
300\,h after any given hard state measurement will not contain even a
single measurement in a different state. This number increases to over
70\% for larger $n$, i.e., when very short excursions to the
soft/intermediate state and/or outliers due to misclassifications are
ignored (Fig.~\ref{fig:cumulative}).

Though less stable and shorter than hard states, soft states generally
show a similar behavior. Even for $n=0$, a significant number of soft
state measurements is not followed by a state transition within
300\,h, i.e., prolonged stable, soft states lasting more than 12\,days
exist. Such states are observed by ASM in periods~\textsc{i}
and~\textsc{iii} (Fig.~\ref{fig:monster}). MAXI and BAT data suggest
an increased occurrence of such stable soft states also in
period~\textsc{v}. 

The intermediate state behaves differently. The transited fraction
strongly increases within the first few hours. Already fifty hours
after a reference intermediate measurement about half of the
measurement will show the source in a different state. The cumulative
transited fraction approaches 100\% even for $n=12$ at 300\,d
(Fig.~\ref{fig:cumulative}).  Intermediate states are therefore short
and unstable compared to hard and soft states.

\section{Summary}\label{sec:summary}

Based on pointed RXTE observations, we provided criteria to define
X-ray states (hard, intermediate, and soft) using light curves from
all sky monitor instruments. In particular we have shown that:
\begin{itemize}
\item due to the complex source behavior, simple state definitions
  based just on the source count rate or just on the hardness do not
  adequately describe states,
\item a combination of RXTE-ASM total count rate and hardness can be
  used to define states before MJD~55200 (see Table~\ref{tap:all_map}
  for exact cut),
\item the best separation of states is achieved with simultaneous ASM
  data. Data within $\Delta t \pm $6\,h result in a contamination of
  $<$10\% for the hard and soft state and $<$30\% for the intermediate
  state,
\item a combination of MAXI count rate and hardness can be used to
  define states (see Table~\ref{tap:all_map} for exact cuts),
\item soft coverage is necessary to define states and the lack of such
  does not allow to distinguish the hard and the intermediate state
  using publicly available BAT light curves. BAT light curves can,
  however, be used to distinguish the soft state from
  hard/intermediate states: the source is in the soft state below a
  threshold of
  $0.09\,\mathrm{counts}\,\mathrm{cm}^{-2}\,\mathrm{s}^{-1}$, and in
  the hard/intermediate state above it (Table~\ref{tap:all_map}),
\item the lack of soft coverage does not allow a separation between
  hard and intermediate state with publicly available GBM data, with
  the analysis being further hindered by the low time resolution of
  the GBM lightcurve, but a rough state classification is to consider
  \mbox{Cyg\,X-1}\xspace in the soft state for fluxes $<$0.6\,Crab and
  in the hard/intermediate state above that threshold
  (Table~\ref{tap:all_map}),
\item the hard state is by far the most stable state of Cyg~X-1,
  followed by the soft state. The probability that the source remains
  in the hard state for at least one week (200\,h) is $>$85\% (using
  $P_{\mathrm{ct},12}$). Soft states are slightly less stable, but the
  probability of a soft state being longer than one week is still
  $\sim$75\%. Intermediate states are short-lived and typically last a
  few days at most, implying that they can only be caught with monitor
  data with a time resolution of better than 1\,d.
\end{itemize}
The state classification introduced here can be reliably used to
define source states where no X-ray continuum spectrum measurement is
available. The high frequency of all sky monitor measurements for
RXTE-ASM, Swift-BAT, and MAXI enables us to catch short flares and
especially quick state transitions.

\begin{acknowledgements} This work has been partially funded by the
  Bundesministerium f\"ur Wirtschaft und Technologie under Deutsches
  Zentrum f\"ur Luft- und Raumfahrt Grants 50\,OR\,1007 and
  50\,OR\,1113 and by the European Commission through ITN 215212
  ``Black Hole Universe'', was partially completed by LLNL under
  Contract DE-AC52-07NA27344, and is supported by NASA grants to LLNL
  and NASA/GSFC. This research has made use of the MAXI data provided
  by RIKEN, JAXA and the MAXI team. We thank John E. Davis for the
  development of the \texttt{slxfig} module used to prepare all
  figures in this work. VG thanks NASA's Goddard Space Flight Center
  for its hospitality during the time when the research presented here
  was done. VG and MCB acknowledge support from
  the Faculty of the European Space Astronomy Centre (ESAC).
\end{acknowledgements}

\bibliographystyle{jwaabib} 
\bibliography{mnemonic,aa_abbrv,references}

\end{document}